\documentclass[prb,aps,floatfix,twocolumn,amsmath,amssymb,showpacs]{revtex4}

\usepackage{epsfig}
\usepackage{bm}
\usepackage{feynmf}
\usepackage{color}
\usepackage{amsmath}
\usepackage{simplewick}
\newcommand{\be}{\begin{equation}}
\newcommand{\ee}{\end{equation}}
\newcommand{\bea}{\begin{eqnarray}}
\newcommand{\eea}{\end{eqnarray}}
\newcommand{\vtr}[1]{\ensuremath{\boldsymbol{#1}}}

\newcommand{\ket}[1]{\ensuremath{\left|{#1}\right\rangle}}
\newcommand{\bra}[1]{\ensuremath{\left\langle{#1}\right|}}
\newcommand{\braket}[1]{\ensuremath{\left\langle{#1}\right\rangle}}

\newcommand{\ocp}[3]{\ensuremath{\frac{n_{#1}+n_{{#1}+Q}-1}{i\omega_n {#2}\epsilon_{#1} {#3}\epsilon_{{#1}+Q}}}}
\newcommand{\ocm}[3]{\ensuremath{\frac{n_{#1}-n_{{#1}+Q}}{i\omega_n {#2}\epsilon_{#1} {#3}\epsilon_{{#1}+Q}}}}
\newcommand{\ocs}[3]{\ensuremath{\frac{n_{q}}{i\omega_n {#2}\epsilon_{#1} {#3}\epsilon_{{#1}+Q}}}}
\newcommand{\ocpw}[3]{\ensuremath{\frac{n_{#1}+n_{{#1}+Q}-1}{i\omega_n {#2}\epsilon_{#1} {#3}\epsilon_{{#1}+Q}}}}
\newcommand{\ocmw}[3]{\ensuremath{\frac{n_{#1}-n_{{#1}+Q}}{i\omega_n {#2}\epsilon_{#1} {#3}\epsilon_{{#1}+Q}}}}

\newcommand{\pairll}[3]{\ensuremath{\alpha_{#1}#3\alpha_{#2}#3}}
\newcommand{\pairrl}[3]{\ensuremath{\alpha_{#1}^{\dagger}#3\alpha_{#2}#3}}
\newcommand{\pairrr}[3]{\ensuremath{\alpha_{#1}^{\dagger}#3\alpha_{#2}^{\dagger}#3}}

\makeatletter
\def\overbracket#1{\mathop{\vbox{\ialign{##\crcr\noalign{\kern3\p@}
\downbracketfill\crcr\noalign{\kern3\p@\nointerlineskip}
$\hfil\displaystyle{#1}\hfil$\crcr}}}\limits}
\def\underbracket#1{\mathop{\vtop{\ialign{##\crcr
$\hfil\displaystyle{#1}\hfil$\crcr\noalign{\kern3\p@\nointerlineskip}
\upbracketfill\crcr\noalign{\kern3\p@}}}}\limits}
\def\overparenthesis#1{\mathop{\vbox{\ialign{##\crcr\noalign{\kern3\p@}
\downparenthfill\crcr\noalign{\kern3\p@\nointerlineskip}
$\hfil\displaystyle{#1}\hfil$\crcr}}}\limits}
\def\underparenthesis#1{\mathop{\vtop{\ialign{##\crcr
$\hfil\displaystyle{#1}\hfil$\crcr\noalign{\kern3\p@\nointerlineskip}
\upparenthfill\crcr\noalign{\kern3\p@}}}}\limits}
\def\downparenthfill{$\m@th\braceld\leaders\vrule\hfill\bracerd$}
\def\upparenthfill{$\m@th\bracelu\leaders\vrule\hfill\braceru$}
\def\upbracketfill{$\m@th\makesm@sh{\llap{\vrule\@height3\p@\@width.7\p@}}%
\leaders\vrule\@height.7\p@\hfill
\makesm@sh{\rlap{\vrule\@height3\p@\@width.7\p@}}$}
\def\downbracketfill{$\m@th
\makesm@sh{\llap{\vrule\@height.7\p@\@depth2.3\p@\@width.7\p@}}%
\leaders\vrule\@height.7\p@\hfill
\makesm@sh{\rlap{\vrule\@height.7\p@\@depth2.3\p@\@width.7\p@}}$}
\makeatother

\def\nn{\nonumber\\}
\def\tt{\tilde{\theta}}

\begin{document}
\DeclareGraphicsExtensions{.eps}
\title{Finite Temperature Dynamical Structure Factor
of the Heisenberg--Ising Chain}
\date{\today}
\author{A. J. A. James$^{1,2}$, W. D. Goetze$^2$ and F. H. L. Essler$^2$}
\affiliation{$^1$Department of Physics, University of Virginia,
  Charlottesville, VA 22904-4717, USA\\
$^2$Rudolf Peierls Centre for Theoretical Physics, University
  of Oxford, Oxford OX1 3NP, UK} 
\begin{abstract}
We consider the spin--1/2 Heisenberg XXZ chain in the regime of large
Ising--like anisotropy $\Delta$. By a combination of duality and
Jordan--Wigner transformations we derive a mapping to weakly
interacting spinless fermions, which represent domain walls between
the two degenerate ground states. We develop a perturbative expansion
in $1/\Delta$ for the transverse dynamical spin structure factor at
finite temperatures and in an applied transverse magnetic field. We
present a unified description for both the low--energy
temperature-activated response and the temperature evolution of the
T=0 two--spinon continuum. We find that the two--spinon continuum {\sl
  narrows} in energy with increasing temperature. At the same time
spectral weight is transferred from the two--spinon continuum to the
low energy intraband scattering continuum, which is strongly peaked
around the position of the (single) spinon dispersion (``Villain mode'').
\end{abstract}

\maketitle
\begin{fmffile}{strong5diags}
\section{Introduction}
The spin--1/2 Heisenberg XXZ chain is a paradigm in
low--dimensional quantum magnetism. Its Hamiltonian is 
\begin{multline}
H= {J}\sum_n \Delta S_n^z S_{n+1}^z +S_{n}^x S_{n+1}^x + S_{n}^y S_{n+1}^y+\sum_n\mathbf{h}\cdot\mathbf{S}_n,
\label{eqnxxzham}
\end{multline}
where $\mathbf{h}$ is an external magnetic field and $\Delta$ controls
the exchange anisotropy. The one--dimensional case is particularly significant because for a field
$\mathbf{h} $ parallel to the $\hat{\mathbf{z}}$ direction, the
Hamiltonian is integrable and the spectrum of the spin chain may be
extracted exactly.\cite{orbach,cloizeaux,yang1,korepin} Furthermore
the Hamiltonian (\ref{eqnxxzham}) is thought to provide a realistic
description of several quasi--1D experimental compounds. Examples
include ${\rm Cs_2CoCl_4}$ for which $\Delta =
0.25$,\cite{algra,duxbury,yoshizawa} ${\rm
  CsCoBr_3}$,\cite{nagler,nagler2} ${\rm CsCoCl_3}$\cite{yoshizawa2}
and ${\rm TlCoCl_3}$\cite{oosawa} all of which have $\Delta \sim 7$. 

For $\Delta>1, h=0$ and $T=0$ the XXZ spin chain is in a N\'{e}el
phase. The fundamental excitations take the form of gapped
fractionalized spin--1/2 quantum solitons known as
spinons.\cite{faddeev} Strictly in the Ising limit, $\Delta=\infty$,
the spinons can be identified simply as domain walls with gap
$\Delta/2$. Experiments have established the existence of several
manifestations of the XXZ model in the Ising
regime.\cite{nagler,nagler2,yoshizawa2,oosawa} In some cases these
experiments have also probed the effects of
temperature\cite{nagler2,braun} and transverse field\cite{braun} on
 dynamical spin--spin correlations. 

The measure of dynamical correlations known as the dynamical structure
factor is an important quantity in the study of quantum magnets.\cite{igor} This is for two reasons: firstly, the dynamical structure
factor is directly measurable by inelastic neutron scattering
experiments and secondly the nature of the dynamical response is
highly specific to the system in question, so that it serves as a
characterization tool. 
A particular feature that one would like to understand in the case of
the XXZ chain is the finite temperature low energy spin response known
as the `Villain mode'.\cite{villain} This response due to scattering
between domain wall pair states has been observed in
Refs. [\onlinecite{nagler2,braun}]. As this response occurs only at
finite temperatures it necessitates a theory that accounts for thermal
fluctuations. 
A recent analysis of the continuum limit of two gapped integrable
quantum spin chains\cite{essrmk} has shown that at raised temperatures
the effect of thermal fluctuations cannot be described in terms of a
simple thermal decoherence or relaxation time picture. Instead markedly
asymmetric thermal broadening of single particle modes is
observed. This paradigm has been found to be in agreement with theoretical
\cite{dimerchain,mikeska} and experimental\cite{tennant1} studies of
the spin--1/2 Heisenberg chain with strongly alternating exchange, a model
which is gapped but not integrable. 
In contrast the gapped excitations in the spin--1/2 XXZ chain occur
only in pairs; it is then interesting to try and understand the
thermal evolution of the resulting two particle response in addition
to that of the Villain mode. 

Despite the advantages afforded by integrability, the task of
calculating correlation functions (and hence dynamics) for the XXZ
chain is still far from simple. First order perturbative treatments
around the $\Delta = \infty$ limit \cite{is,nagler,villain} and
$1/S$ expansions \cite{mikeska2} have given some insight.
In recent years significant progress has been made for the $\Delta >
1$ regime, both via Bethe's Ansatz\cite{isaac} and a different exact
technique which works directly with the thermodynamic
limit.\cite{jimbobook} This has lead to an exact expression for the
transverse dynamical correlations at $T=0$.\cite{bg,isaac} Results at
finite temperature are generally still limited to asymptotically exact
thermodynamic
quantities\cite{thermo,fabsbook,takahashibook,johnson1}. However there 
have been promising advances that rely on generalizing multiple integral
representations for time dependent correlation functions to finite
temperatures.\cite{kitanine,gohmann1,gohmann2,sakai} Currently these
methods have not yielded expressions for the most experimentally
relevant quantity, the dynamical structure factor. 

In this paper we present a perturbative calculation for the transverse
spin response in the $\Delta \gg 1$ limit, valid at finite
temperatures and capable of incorporating a transverse field. We note
that the temperature dependence of the dynamical structure factor in
the critical $-1<\Delta\leq 1$ regime has been determined by exact
diagonalization of short chains and very recently by Quantum Monte
Carlo and DMRG computations.\cite{finiteTXXZ} The
usual perturbative approach to the XXZ chain uses the Jordan--Wigner
transformation\cite{jw} to produce an expansion in powers of
$\Delta$. This is suitable for investigating the XY, $\lvert \Delta
\rvert \ll 1$ case but inadequate here. Instead we first perform a
Kramers--Wannier\cite{kw} duality transformation to a new Hamiltonian
in terms of dual operators. A Jordan--Wigner transformation from these
dual operators to spinless fermions then leads to a controlled
expansion in $1/\Delta$.  An equivalent mapping has been used
previously, coupled with mean field theory, to find the approximate
excitation spectrum in the Ising phase.\cite{gs} We take an
alternative approach resumming certain terms in the expansion to all
orders, in order to take account of both quantum and thermal
fluctuations. 

The structure of the paper is as follows. First in
Sec. \ref{xxzsym} we discuss symmetries of the Hamiltonian and the
dynamical structure factor. Second, in
Sec. \ref{secxxztrans} we transform the Hamiltonian into a form
suitable for the expansion. In Sec. \ref{secxxzdyn} we describe
the perturbative expansion of the transverse spin--spin correlator. In
Sec. \ref{secxxzresum} we explain how to resum certain terms in this
expansion in order to obtain finite results. In
Sec. \ref{secxxzresults} we discuss the behaviour of the dynamical
structure factor for a range of parameters. Sec. \ref{xxzconclusion}
contains some brief concluding remarks. 
\section{Symmetries of the Hamiltonian and the Structure Factor}
\label{xxzsym}
We now describe the symmetries of the Hamiltonian and their
consequences for spin--spin correlation functions. For $h=0$
the Hamiltonian, Eq. (\ref{eqnxxzham}), is invariant
under arbitrary rotations $\mathcal{R}^z(\phi)$ around the z-axis as
well as under rotations by $\pi$ around the $x$ axis,
$\mathcal{R}^x(\pi)$, which entail the mapping 
\begin{align*}
S_j^x &\to S_j^x, \nn
S_j^{y,z} &\to -S_j^{y,z}.
\end{align*}
The two types of symmetry operations do not commute, but we can
diagonalize the Hamiltonian simultaneously with either $S^z$ or with
the generator $\mathcal{R}^x(\pi)$ of the $\mathbb{Z}_2$ symmetry.
This in turn implies that all off--diagonal spin correlators vanish for
$T>0$, by the following arguments. When considering the thermal expectation values $\langle S_n^a S_m^z\rangle$ with $a=x,y$ we choose a basis of energy
eigenstates in which the total $S^z$ is diagonal. Then carrying out a
rotation by $\pi$ around the z-axis sends $S^a_n\rightarrow -S^a_n$, $a=x,y$
and as a result
\begin{align}
\langle S_n^a S_m^z \rangle  = -\langle S_n^a S_m^z \rangle & = 
0\ ,\ a=x,y.
\end{align}
On the other hand, when considering $\langle S_n^x S_m^y \rangle$ we
use a basis of simultaneous eigenstates of $H$ and
$\mathcal{R}^x(\pi)$ to carry out the thermal trace. Under the
$\mathbb{Z}_2$ symmetry the thermal expectation value is negated, leading to
\begin{align}
\langle S_n^x S_m^y \rangle = -\langle S_n^x S_m^y \rangle = 0.
\end{align}
This shows that in the absence of the transverse magnetic field all
off--diagonal elements of the dynamical structure factor vanish.
In the presence of a finite transverse field we only have the
$\mathbb{Z}_2$ to work with. Concomitantly for $h>0$ one finds
$\langle S_n^x  S_m^{y} \rangle=\langle S_n^x S_m^{z} \rangle=0$ but
$\langle S_n^y S_m^z \rangle$ is no longer required to vanish by
symmetry and as a result acquires a finite value.
\section{Transformations of the Hamiltonian}
\label{secxxztrans}
In order to proceed we aim to re--express the Hamiltonian in terms of
spinless fermions by means of a Jordan--Wigner transformation in such a
way that we analyze the resulting interacting fermion Hamiltonian by
standard perturbative methods. In order to achieve this, the 
$\Delta J\sum_nS^z_nS^z_{n+1}$ part
of the Hamiltonian (\ref{eqnxxzham}) must be mapped to an expression
quadratic in fermions. One approach for doing this is outlined in
Appendix \ref{sec:direct}, another one is discussed in detail next.
In the following we consider a {\sl transverse} magnetic field applied
along the $\hat{\mathbf{x}}$ direction. We note that the effects of of
a transverse field in the critical region of the XXZ--chain
$-1<\Delta\leq 1$ have been studied in some detail in Refs
\onlinecite{cs2cucl4}. 
\subsection{Duality Transformation}
We work on the infinite chain so that we can ignore boundary
effects. The Hamiltonian, Eq. (\ref{eqnxxzham}), in terms of Pauli
spin matrices is given by  
\begin{align}
&H=H_\Delta+H_h,\nn
&H_\Delta=\frac{J}{4} \Delta \sum_n \sigma_n^z \sigma_{n+1}^z + \frac{J}{4} \sum_n \big(\sigma_{n}^x \sigma_{n+1}^x + \sigma_{n}^y \sigma_{n+1}^y \big),\nn
&H_h=\frac{h}{2}\sum_n\sigma_n^x.
\end{align}
The Kramers--Wannier duality transformation is defined by
\bea
\mu_{n+1/2}^{x}&=&\sigma_{n}^z \sigma_{n+1}^z ,\quad
\mu_{n+1/2}^{z}=\prod_{j<n+1} \sigma_j^x,\nn
\mu_{n+1/2}^y&=&-i \mu_{n+1/2}^z \mu_{n+1/2}^x. 
\label{eqnxxzkw}
\eea
This transformation defines operators on a dual lattice and maps an
Ising chain from its ordered to its disordered phase.\cite{kogut} 
Applying the transformation to the XXZ Hamiltonian we find
\begin{align}
\begin{split}
&H_\Delta=\frac{J}{4} \sum_n\big(\Delta \mu_{n+1/2}^x \\
&\qquad+ \mu_{n-1/2}^{z} \mu_{n+3/2}^{z}- \mu_{n+1/2}^{x} \mu_{n-1/2}^{z} \mu_{n+3/2}^{z} \big),
\end{split}
\nn
&H_h=\frac{h}{2}\sum_n\mu_{n-1/2}^{z} \mu_{n+1/2}^{z}.
\label{eqnxxzkwham}
\end{align}
We note that applying the duality transformation to a finite
open chain leads to a dual Hamiltonian containing additional boundary
terms that in particular ensure a doubly degenerate ground state in
the thermodynamic limit. This is most easily seen for $\Delta \gg h \gg 1$, i.e. the Ising model in a transverse field. In this limit the mapping gives
\begin{align}
&\Delta\sum_{n=1}^{N-1}\sigma_n^z \sigma_{n+1}^z +h \sum_{n=1}^{N}\sigma_n^x \nn
&\to h\sum_{n=1}^{N-1}\mu_{n+1/2}^z \mu_{n+3/2}^z +\Delta \sum_{n=1}^{N-1}\mu_{n+1/2}^x
+h\mu_{3/2}^z\ .
\end{align}
Now as $h/\Delta \to 0$ the two ground states of the
original Hamiltonian become the familiar N\'{e}el states,
$\big\langle\sigma^z_n\big\rangle=-\big\langle\sigma^z_{n+1}\big\rangle $.
In contrast, for the dual Hamiltonian the ground state is
given by $\big\langle\mu_{n+1/2}^x\big\rangle=-1$ with $n<N$. The twofold
degeneracy is then maintained by the $N$th dual spin, $\mu_{N+1/2}$
which is free to point in either direction. 

In the following we will be interested only in bulk correlations of operators that are local under the duality
transformation. Hence the boundary terms do not play a role and will be
dropped.
\subsection{Fermionic Representation}
In order to proceed further it is necessary to map the spins to fermions.
We perform a rotation of spin axes
\begin{align}
\begin{array}{cc}
\mu_{n+1/2}^x \rightarrow \tau_n^z , & \mu_{n+1/2}^z \rightarrow \tau_n^y ,
\end{array}
\end{align}
with raising and lowering operators
\begin{align}
\begin{array}{cc}
\tau_n^+=\frac{\tau_n^x+i\tau_n^y}{2} , & \tau_n^-=\frac{\tau_n^x-i\tau_n^y}{2}
\end{array}
\end{align}
and then use the Jordan--Wigner transformation:
\begin{align}
\tau_{n}^z & = 2 c^\dagger_n c_n -1, \nn
\tau_{n}^+ & = c_n^\dagger e^{-i \pi \sum_{j<n} c^\dagger_j c_j}.
\end{align}
\subsubsection{Spin Operators}
We first consider the transformation of the lattice spin operators
under the mappings. Crucially, the transverse spin operator is local
under the transformations
\begin{align}
\sigma^x_n=c^\dagger_{n-1}c_n-c^\dagger_{n-1}c^\dagger_n+{\rm h.c.}\ .
\end{align}
On the other hand, both $\sigma^z$ and $\sigma^y$ acquire 
Jordan-Wigner strings. As a result, our formalism will allow us to
determine the $xx$--component of the dynamical structure factor
only. It follows from the symmetry considerations above that in
absence of a transverse magnetic field this suffices to determine all
transverse correlations. 
\subsubsection{Hamiltonian}
After the Jordan--Wigner transformation the Hamiltonian takes the form
\begin{widetext}
\begin{align}
H_\Delta = \frac{J}{2} \Delta \sum_n c^\dagger_n c_n+ \frac{J}{2}
\sum_n \big[ c^\dagger_{n-1} c_{n+1} - c^\dagger_{n-1}
  c^\dagger_{n+1}- c^\dagger_{n-1} c_n^\dagger c_n c_{n+1} +
  c^\dagger_{n-1} c_n^\dagger c_n c^\dagger_{n+1} + \textrm{h.c.}
  \big] +{\rm const}.
\end{align}
\end{widetext}
and
\begin{align}
H_h = \frac{h}{2}\sum_n \left( c_{n-1}^\dagger c_{n} - c_{n-1}^\dagger c_{n}^\dagger + \textrm{h.c.} \right).
\end{align}
We now write the Hamiltonian as a sum of two pieces, $H=H_2+H_4$, containing terms quadratic and quartic in the fermionic operators respectively.
The quartic (interaction) terms are $\mathcal{O}(\Delta^0)$ while the quadratic pieces mix orders $\mathcal{O}(\Delta)$ and $\mathcal{O}(\Delta^0)$. The external field only appears in the quadratic part of the Hamiltonian.
After taking the Fourier transform of the quadratic part, $H_2$, of the Hamiltonian we find
\begin{align}
H_2 & =  \frac{J}{8} \sum_k \begin{pmatrix} c_k^\dagger & c_{-k} \end{pmatrix} \begin{pmatrix}A_k & iB_k \\-iB_k & -A_k\end{pmatrix} \begin{pmatrix}c_k \\c_{-k}^\dagger \end{pmatrix}
\end{align}
with  $A_k=2\Delta+4\cos(2k)+\frac{4h}{J}\cos(k)$ and $B_k=4 \sin(2k)+\frac{4h}{J}\sin(k)$.
This can be diagonalized by a Bogoliubov transformation of the form
\begin{align}
\begin{pmatrix}c_k\\c^\dagger_{-k}\end{pmatrix} & = \begin{pmatrix}i \cos(\theta_k)&-\sin(\theta_k)\\ \sin(\theta_k)&-i\cos(\theta_k)\end{pmatrix} \begin{pmatrix}\alpha_k \\ \alpha_{-k}^\dagger\end{pmatrix}
\label{bogtrafo}
\end{align}
so that
\begin{align}
H_2 & = \frac{1}{2} \sum_k \begin{pmatrix} \alpha_k^\dagger & \alpha_{-k} \end{pmatrix} \begin{pmatrix}\epsilon_k &0\\0&-\epsilon_k\end{pmatrix} \begin{pmatrix}\alpha_k\\\alpha_{-k}^\dagger\end{pmatrix}
\label{eqnxxzH2}
\end{align}
with $\tan(2\theta_k)=B_k/A_k$ and $\epsilon_k=\frac{J}{4} \sqrt{A_k^2+B_k^2}$.

Now we consider the quartic part of the Hamiltonian:
\begin{align}
H_4 & = -\frac{J}{2} \sum_n \left( c^\dagger_{n-1} c_n^\dagger c_n c_{n+1} + c^\dagger_{n-1} c_n^\dagger  c^\dagger_{n+1} c_n + \textrm{h.c.}  \right).
\end{align}
Taking the Fourier transform and manipulating indices leads to
\begin{align}
H_4  =& -\frac{J}{4N}\sum_{k_1,k_2,k_3,k_4} \delta_{k_1+k_2+k_3+k_4} \nn
&\quad \times\big[f_{(k_1k_2)(k_3k_4)}c^\dagger_{k_1} c_{k_2}^\dagger c_{-k_3} c_{-k_4}\nn
&\qquad+\frac{2}{3}\left(ig_{(k_1k_2k_3)}c^\dagger_{k_1} c_{k_2}^\dagger  c^\dagger_{k_3} c_{-k_4} +\textrm{h.c.} \right) \big],
\end{align}
where we have defined the new functions
\begin{align}
\begin{split}
f_{(k_1,k_2)(k_3,k_4)} & =\cos(k_1-k_4)-\cos(k_2-k_4)\\ &\quad-\cos(k_1-k_3)+\cos(k_2-k_3),
\end{split} \\\nn
\begin{split}
g_{(k_1,k_2,k_3)} & = \sin(k_1-k_2)+\sin(k_2-k_3)\\
&\quad+\sin(k_3-k_1).
\end{split}
\end{align}
The function $f$ is antisymmetric under exchange of the two momenta
within a pair of brackets. The function $g$ is symmetric for cyclic
permutations of its momentum arguments and antisymmetric
otherwise. For example
$f_{(k_1,k_2)(k_3,k_4)}=-f_{(k_2,k_1)(k_3,k_4)}=f_{(k_3,k_4)(k_1,k_2)}$
and
$g_{(k_1,k_2,k_3)}=g_{(k_2,k_3,k_1)}=-g_{(k_2,k_1,k_3)}$. Performing
the Bogoliubov transformation on $H_4$ is standard but lengthy. By
manipulating indices under the sums we arrive at a relatively compact
form for the part quartic in Bogoliubov operators:
\begin{widetext}
\begin{align}
H_4=&\frac{1}{N}\sum_{1,2,3,4}\delta_{k_1+k_2+k_3+k_4,0}
\Bigl\{V_0(k_1,k_2,k_3,k_4)\left[\alpha^{\dagger}_{k_1}\alpha^{\dagger}_{k_2}
\alpha^{\dagger}_{k_3}\alpha^{\dagger}_{k_4}+{\rm h.c.}\right]\nn
&\qquad\qquad+\left[V_1(k_1,k_2,k_3,k_4)\alpha^{\dagger}_{k_1}\alpha_{-k_2}
\alpha_{-k_3}\alpha_{-k_4}+{\rm h.c.}\right]
+V_2(k_1,k_2,k_3,k_4)
\alpha^{\dagger}_{k_1}\alpha^{\dagger}_{k_2}\alpha_{-k_3}\alpha_{-k_4}
\Bigr\}.
\label{eqnxxzH4}
\end{align}
The interaction vertices are given by
\begin{align}
&V_0(k_1,k_2,k_3,k_4)=\frac{J}{96} \sum_{P \in S_4} \mathrm{sgn}(P) \cos(k_{P(1)}-k_{P(2)}-\theta_{k_{P(1)}}+\theta_{k_{P(2)}}+\theta_{k_{P(3)}}-\theta_{k_{P(4)}}),
\end{align}
with permutation $P$ acting on the set $\{1,2,3,4\}$,
\begin{align}
V_1(k_1,k_2,k_3,k_4)=&i\frac{J}{12}\sum_{P \in S_3} \mathrm{sgn}(P)
[\sin(k_{1}-k_{P(2)}-\theta_{k_{1}}+\theta_{k_{P(2)}}+\theta_{k_{P(3)}}-\theta_{k_{P(4)}})
  \nn 
&\qquad -\sin(k_{P(2)}-k_{P(3)}+\theta_{k_{1}}-\theta_{k_{P(2)}}+\theta_{k_{P(3)}}-\theta_{k_{P(4)}})],
\end{align}
with permutation $P$ acting on the set $\{2,3,4\}$ and finally
\begin{align}
V_2(k_1,k_2,k_3,k_4)=\frac{J}{8}\Big(&\sum_{P \in S_3}\mathrm{sgn}(P)\cos(k_{3}-k_{4}-\theta_{k_{3}}+\theta_{k_{4}}-\theta_{k_{P(1)}}+\theta_{k_{P(2)}})\nn
&+\sum_{P' \in S_3}\mathrm{sgn}(P')\cos(k_{1}-k_{2}-\theta_{k_{1}}+\theta_{k_{2}}-\theta_{k_{P'(3)}}+\theta_{k_{P'(4)}})\nn
&+\sum_{P \in S_3}\sum_{P' \in S_3}\mathrm{sgn}(P)\mathrm{sgn}(P')\big[\cos(k_{P(1)}-k_{P'(3)}-\theta_{k_{P(1)}}+\theta_{k_{P(2)}}+\theta_{k_{P'(3)}}-\theta_{k_{P'(4)}})\nn
&+\cos(k_{P(1)}-k_{P'(3)}-\theta_{k_{P(1)}}-\theta_{k_{P(2)}}+\theta_{k_{P'(3)}}+\theta_{k_{P'(4)}})\big]\Big),
\end{align}
where $P$ and $P'$ act on $\{1,2\}$ and $\{3,4\}$ respectively.
\end{widetext}

New quadratic terms are generated by normal ordering the quartic
piece. We must then include these with the original terms from $H_2$
and solve for $\theta_k$ self--consistently so that the off--diagonal
terms are zero. This requirement may be recast as a self--consistency
condition for every $k$: 
\begin{align}
\tan(2\theta_k)=\frac{2\sin(2k)+\frac{2h}{J}\sin(k)+\frac{1}{2N}\sum_q \Theta_2(k,q)}{
\Delta+2\cos(2k)+\frac{2h}{J}\cos(k)+\frac{1}{N}\sum_q\Theta_1(k,q)},
\label{eqnselfcons}
\end{align}
where we have defined
\begin{align}
 \Theta_1(k,q) & = 2f_{(k,q)(-k,-q)} \sin^2(\theta_q)+g_{(k,q,-q)} \sin(2\theta_q), \\
 \Theta_2(k,q) & = f_{(k,-k)(q,-q)} \sin(2\theta_q)-4g_{(k,-k,q)}\sin^2(\theta_q).
\end{align}
\begin{figure}[t]
\begin{center}
\epsfxsize=0.45\textwidth
\epsfbox{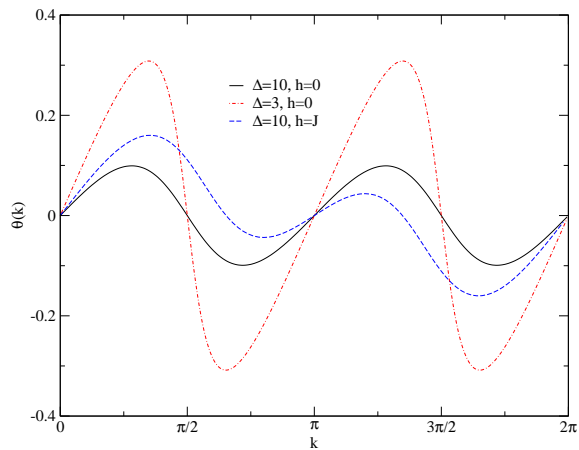}
\caption{(Color Online) The self--consistent Bogoliubov parameter $\theta_k$, for $J=1$. The parameter scales as $\Delta^{-1}$.} 
\label{figxxztheta}
\end{center}
\end{figure}
\begin{figure}[t]
\begin{center}
\epsfxsize=0.45\textwidth
\epsfbox{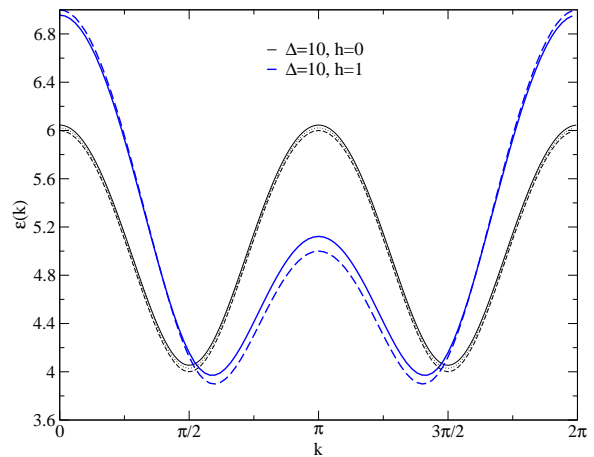}
\caption{(Color Online) The single particle dispersion relation, $\epsilon_k$ (with $J=1$) for $\Delta=10$. The dispersion calculated with and without self--consistency (solid and dashed curves respectively) and the exact spinon dispersion\cite{bg,isaac} (dotted curves) are shown. The spinon result is not available in the case of a finite transverse field. For $h=0$ the dispersion is nearly sinusoidal. Note the functions are $\pi$ periodic for $h=0$ and $2\pi$ periodic otherwise.} 
\label{figxxzdispDelta10}
\end{center}
\end{figure}
\begin{figure}[t]
\begin{center}
\epsfxsize=0.45\textwidth
\epsfbox{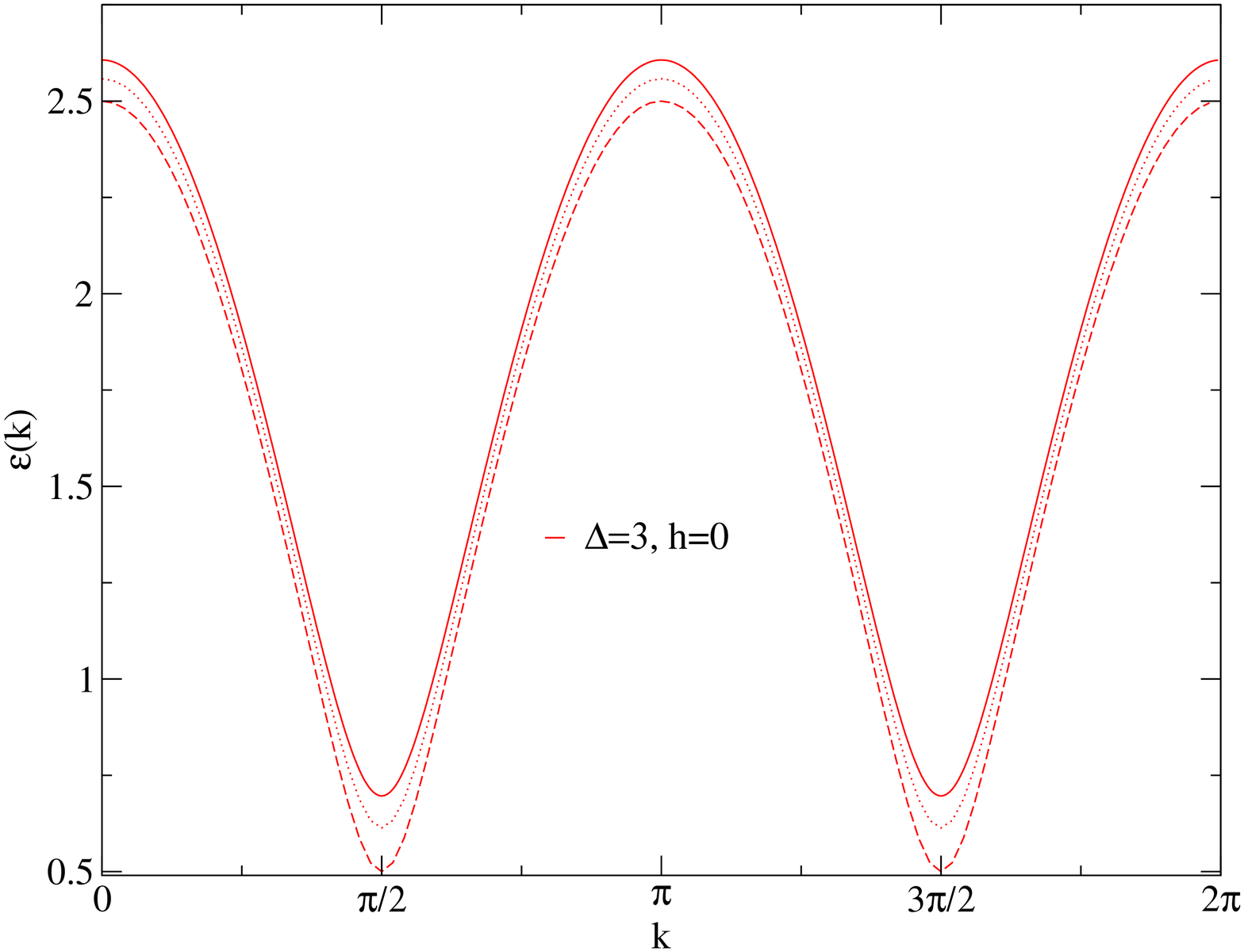}
\caption{(Color Online) The single particle dispersion relation, $\epsilon_k$ (with $J=1$) for $\Delta=3$. The dispersion calculated with and without self--consistency (solid and dashed curves respectively) and the exact spinon dispersion\cite{bg,isaac} (dotted curves) are shown.} 
\label{figxxzdispDelta3}
\end{center}
\end{figure}
Clearly the dispersion is also affected by the new quadratic parts, becoming
\begin{align}
\epsilon_k=J &\Big(\Big[\frac{\Delta}{2}+\cos(2k)+\frac{h}{J}\cos(k)+\frac{1}{2N}\sum_q \Theta_1(k,q)\Big]^2\nn
&+\Big[ \sin(2k)+\frac{h}{J}\sin(k)+\frac{1}{4N}\sum_q \Theta_2(k,q) \Big]^2  \Big)^\frac{1}{2}.
\label{eqnselfdisp}
\end{align}
Evaluating the self--consistency and dispersion relations (\ref{eqnselfcons},\ref{eqnselfdisp}) numerically we can compare the gap (i.e. lowest excitation energy) to the mean field result found by G\'{o}mez--Santos.\cite{gs} Summing over 100 sites our results for the physical (two particle) gap are in excellent agreement. The self--consistent Bogoliubov parameter is plotted in Fig. \ref{figxxztheta} for a range of parameters. Figures \ref{figxxzdispDelta10} and \ref{figxxzdispDelta3} show that $\epsilon_k$ is an excellent approximation to the spinon dispersion. It is also apparent that use of a self--consistent Bogoliubov parameter is a small effect on the level of the dispersion, except in the presence of a transverse field.

It is worth emphasising that the fermions that feature in the diagonalized quadratic Hamiltonian, Eq. (\ref{eqnxxzH2}), are not the same as the spinons of the exact treatment.\cite{jimbobook,bg,isaac} Here fermion number is not conserved by the interaction vertices. In contrast spinon number is conserved by the exact solution. Instead the fermions described by $\alpha^\dagger_k$ should be viewed as propagating domain walls.\
\subsubsection{Properties of the Eigenstates}
Previously it has been suggested that the fundamental excitations of Heisenberg--Ising chains are chiral.\cite{braun} We now make some remarks on this possibility, in light of our results.
The relevant chiral operator, $C_x$, is defined as
\be
\label{conservedchirality}
C_x  = \hat{\vtr{x}}\cdot\sum_n \vtr{S}_n\times \vtr{S}_{n+1}
 = \sum_n S_n^y S_{n+1}^z - S_n^z S_{n+1}^y.
\ee
For chirality to be a good quantum number for the XXZ chain, $C_x$ must
commute with the Hamiltonian. We find
\bea
[H,C_x] &=& \sum_{n}  i(S^x_{n-1}-S^x_{n}+S^x_{n+1}-S^x_{n+2})\nn
&&\times(S_n^yS_{n+1}^y+S_n^zS_{n+1}^z)\ ,
\eea
which is an $\mathcal{O}(1)$ contribution. A priori this
demonstrates that at least one eigenstate of the Hamiltonian is not an
eigenstate of $C_x$. However it is possible to show that spinon states
are generally not chirality eigenstates. First we write the commutator
in the fermionic basis 
\bea
[H,C_x] &=& -\frac{i}{4}\sum_{n}
\big( c_n^\dagger c_n-\frac{M_{n-1,n+1}}{2}+c_n^\dagger c_n M_{n-1,n+1} \big)\nn
&&{\hskip -40pt}\times \big(
M_{n-2,n-1}-M_{n-1,n}+M_{n,n+1}-M_{n+1,n+2}\big),
\label{conservedchiralityf}
\eea
where we have defined
\begin{align}
M_{n,n+1}=c_n^\dagger c_{n+1}^\dagger+c_n c_{n+1}-c_n^\dagger c_{n+1}-c_n c_{n+1}^\dagger.
\end{align}
As written, Eq. (\ref{conservedchiralityf}) contains only terms
quartic and sextic in the creation and annihilation operators. However
once Fourier transformed, rewritten in the Bogoliubov basis
($\alpha_k^\dagger$) and normal ordered, quadratic terms will be
generated. A true (one--spinon) excitation of the system, $\ket{\Psi}$,
involves a superposition of domain walls created by the
$\alpha_k^\dagger$ operators. Schematically 
\begin{align}
\label{schematicspinon}
\ket{\Psi} =
\sum_{i=1}^{N}\prod_{k_1,\cdots,k_i}l_i(k_1,\cdots,k_i)\Delta^{1-i}\alpha_{k_i}^\dagger
\ket{0}. 
\end{align}
where the $l_i(k_1,\cdots,k_i)$ are $c$--number functions of the
momenta and we have written the small expansion parameter
$\Delta^{1-i}$ explicitly. It has been shown that at lowest order the
excitations are eigenstates of the chirality
operator.\cite{braun} However from Eqs. (\ref{conservedchiralityf})
and (\ref{schematicspinon}) we see that the expectation
$\bra{\Psi} [H,C_x] \ket{\Psi}$ will generally not be zero, instead
having a finite contribution at lower orders in $\Delta^{-1}$. As
$\Delta \to \infty$ these contributions vanish and at the Ising point
the excitations can be chosen as chirality eigenstates.
\section{Dynamical Response}
\label{secxxzdyn}
The quantity of interest for inelastic neutron scattering is the dynamical structure factor $S^{ab}(\omega, Q)$ given by
\begin{align}
S^{ab}(\omega, Q)&=\frac{1}{N}\int_{-\infty}^{\infty}\frac{dt}{2\pi}\sum_{l,l'}e^{i\omega t}e^{-iQ(l-l')}\langle S^a_l(t)S^b_{l'} \rangle.
\end{align}
Here $S^a_l=\frac{1}{2}\sigma^a_l$ is the $a$ component of the spin
operator at site $l$. 
The structure factor is related to the retarded dynamical
susceptibility $\chi_R^{ab}(\omega, Q)$
\begin{align}
S^{ab}(\omega, Q)&=-\frac{1}{\pi}\frac{1}{1-e^{-\beta\omega}}
{\rm Im}\left[\chi_R^{ab}(\omega, Q)\right],
\end{align}
where
\begin{align}
\chi^{ab}_R(\omega, Q)&=\int_0^\beta d\tau\ e^{i\omega_n\tau}\chi^{ab}(\tau,Q) \biggr|_{\omega_n\rightarrow \eta-i\omega},\nonumber \\
\chi^{ab}(\tau,Q)&=-\frac{1}{N}\sum_{l,l'}e^{-iQ(l-l')}\langle T_\tau S^a_l(\tau)S^b_{l'}\rangle.
\label{eqnxxzsuscept}
\end{align}
Here we have introduced the Matsubara formalism and the expectation
implies a thermal trace $\langle \cdots \rangle = Z^{-1} \sum_m
\langle m \lvert e^{-\beta H} \cdots \rvert m \rangle$. 

Following the discussion in Sec. \ref{xxzsym} it is apparent that for
$h=0$, $\chi^{ab}(\omega, Q)$ and hence $S^{ab}(\omega, Q)$ will be
diagonal in the indices $a,b$ but that this is no longer the case for
$h>0$, in agreement with experiment.\cite{braun}  
\subsection{Dynamical Structure Factor in the Fermionic Representation}
As previously discussed, the Jordan--Wigner transformation introduces non--local `strings' which make calculating $\chi^{zz}$ complicated. We instead focus our attention on the transverse susceptibility, $\chi^{xx}$. 

First the required spin operator must be written in terms of the new
fermionic operators: 
\begin{multline}
\sigma_Q^x= \frac{1}{\sqrt{N}} \sum_k e^{i\frac{Q}{2}} \big[2\cos(k-\theta_k-\theta_{k+Q}+Q/2) \alpha_k^\dagger \alpha_{k+Q} \\ -i \sin(k-\theta_k-\theta_{k+Q}+Q/2)(\alpha_k^\dagger \alpha_{-k-Q}^\dagger - \alpha_{-k} \alpha_{k+Q})\big].\label{eqnxxzspinop}
\end{multline}
We will use Eq. (\ref{eqnxxzspinop}) to evaluate the time ordered dynamical susceptibility,
\begin{align}
\chi^{xx}(\tau,Q) =- \frac{1}{4}\braket{T_{\tau} \sigma_Q^x(\tau) \sigma_{-Q}^x}.
\label{fullcorrelator}
\end{align}
As we aim to calculate (\ref{fullcorrelator}) in perturbation theory in
$H_4$ we now switch to the interaction picture. In order to simplify
the perturbative calculation of $\chi^{xx}$ it is useful to express
(\ref{fullcorrelator}) in terms of a $3\times 3$ matrix
$\Pi_{\beta\gamma}(\tau,Q|k,k')$ (the matrix indices take values
$\beta,\gamma=1,2,3$) as follows 
\begin{align}
 \chi^{xx}(\tau,Q)&=\frac{1}{N^2}\sum_{k,k'} L_\beta(k)
 \Pi_{\beta\gamma}(\tau,Q|k,k')L^\dagger_\gamma(k'), 
\label{chiPi}
\end{align}
where
\begin{align}
L_\beta(k) = & ( \begin{array}{ccc}\frac{i}{2} \sin(\gamma_k), &
  \cos(\gamma_k), & -\frac{i}{2} \sin(\gamma_k) \end{array} )_\beta, \\ 
\gamma_k = & k+Q/2 - \theta_k -\theta_{k+Q},
\end{align}
and the $3\times 3$ matrix $\Pi_{\beta\gamma}(\tau,Q|k,k')$ is given by
\be
\Pi_{\beta\gamma}(\tau,Q|k,q)  =-
\bigl\langle T_\tau X_{\beta\beta}(\tau,Q|k)\
X^\dagger_{\gamma\gamma}(0,Q|q) U(\beta) 
\bigr\rangle, \label{eqnchiU}
\ee
\be
X_{\beta\nu}(\tau,Q|k) =
 \begin{pmatrix}
\pairll{-k}{k+Q}{} & 0 &0 \\
0&\pairrl{k}{k+Q}{} &0\\
0&0&\pairrr{k}{-k-Q}{}
\end{pmatrix}_{\beta\nu}.
\ee
The imaginary time evolution operator in the interaction picture is
\begin{align}
U(\tau) = & T_\tau \exp \left( -\int_{0}^{\tau}d\tau_1 H_4(\tau_1) \right).
\end{align}
The Fourier transform of the matrix $\boldsymbol\Pi$ is given by
\bea
\boldsymbol{\Pi}(i\omega_n,Q|k,q) & =&  
\int_0^\beta d\tau e^{i\omega_n \tau}\ \Pi(\tau,Q|k,q). 
\eea
\subsection{Zeroth Order}
At zeroth order in perturbation theory we replace $U(\beta)$ in
(\ref{eqnchiU}) by $1$. All off--diagonal elements then vanish and we
find ($\omega_{nm}=\omega_n-\omega_m$)
\begin{widetext}
\bea
\Pi^0_{11}(i\omega_n,Q|k,q) &=&\frac{1}{\beta}
\sum_{i\omega_m}G_0(i\omega_{nm},k+Q)G_0(i\omega_m,-k)\left[
\delta_{k,-q-Q}-\delta_{k,q}\right] =
\ocpw{k}{-}{-}\left[\delta_{k,-q-Q}-\delta_{k,q}\right], \\
\Pi^0_{22}(i\omega_n,Q|k,q)  & =& \frac{1}{\beta}\sum_{i\omega_m}
G_0(i\omega_m,k+Q)G_0(-i\omega_{nm},k)\delta_{k,q}  = \ocmw{k}{+}{-}
\; \delta_{k,q}, \\
\Pi^0_{33}(i\omega_n,Q|k,q) &
=&\frac{1}{\beta}\sum_{ik_n}G_0(-i\omega_{nm},-k-Q)G_0(-i\omega_m,k)
\left[\delta_{k,-q-Q}-\delta_{k,q}\right]
=\ocpw{k}{+}{+}\left[\delta_{k,q}-\delta_{k,-q-Q}\right].\nn
\eea
\end{widetext}
Here $n_k=1/(e^{\beta\epsilon_k}+1)$  and the bare Green's function is
given by  
\begin{align}
G_0(ik_n,k)=\frac{1}{ik_n-\epsilon_k}.
\end{align}
The dynamical susceptibility at zeroth order in perturbation theory is
then obtained by substituting the matrix $\Pi^0$  into (\ref{chiPi})
and carrying out the momentum sums. Taking the thermodynamic limit
and analytically continuing to real frequencies, $i\omega_n
\rightarrow \omega+i\eta$, we arrive at the following expression for
the zeroth order retarded susceptibility
\begin{widetext}
\begin{multline}
\chi^{xx}_{R,0}(\omega,Q) =
-\int^{\pi}_{-\pi} \frac{dk}{8\pi} \bigg[(1- \cos(2k+Q -2[\theta_k +
    \theta_{k+Q}]))
\left(\frac{n_k+n_{k+Q}-1}{\omega+i\eta-\epsilon_k-\epsilon_{k+Q}} +
\frac{1-n_k-n_{k+Q}}{\omega+i\eta+\epsilon_k+\epsilon_{k+Q}} \right)
\\ 
 -2(1+\cos(2k+Q -2(\theta_k + \theta_{k+Q})))
  \frac{n_{k+Q}-n_k}{\omega+i\eta-\epsilon_k+\epsilon_{k+Q}} \bigg]. 
\end{multline}
\end{widetext}
The remaining $k$--integral cannot be carried out analytically in
general as the Bogoliubov parameters $\theta_k$ need to be determined
self--consistently and are therefore only known implicitly. However,
in the limit $\Delta\to\infty$ the integral can be taken and simple
expressions for $\chi^{xx}_{R,0}(\omega,Q)$ may be obtained. 
\subsubsection{The $\Delta \to \infty, T=0$ Limit}
In order to evaluate the susceptibility further we take $h=0$ and
expand $\epsilon_k$ as a series in $1/\Delta$. This gives
\be
\epsilon_k =\frac{J\Delta}{2} + J \cos(2k) + \frac{J}{\Delta}
\sin(2k) + \hdots,
\ee
\bea
\epsilon_{k+Q}+\epsilon_k &=&J(\Delta+2\cos(2k+Q)\cos(Q))+\ldots,\\
\epsilon_{k+Q}-\epsilon_k &=&-2J \sin(2k+Q) \sin(Q)+\ldots.
\eea
We see that poles occur at
\begin{align}
\omega & = 2J \sin(2k_0+Q)\sin(Q), \\
\omega & =J(\Delta+2 \cos(2k_{-}+Q)\cos(Q)), \\
\omega & = -J(\Delta+2 \cos(2k_{+}+Q)\cos(Q)).
\end{align}
The contribution from $\theta_k$ only enters at $O(\frac{1}{\Delta})$
so we neglect it. We wish to take the imaginary part of the retarded
susceptibility as this is proportional to the dynamical structure
factor. Defining
\begin{align}
P(\omega,E) & = \Theta(|E|+\omega)\Theta(|E|-\omega) = \left\{ \begin{array}{ll}
0 & \mbox{if $|\omega|>|E|$} \\
1 & \mbox{if $|\omega| \le |E|$}
\end{array} \right.
\end{align}
where the $\Theta$'s are Heaviside step functions, we find
\begin{align}
&-\mathrm{Im}\chi^{xx}_{R,0} (Q,\omega)\approx \nn & (1+\cos(2k_0+Q))\frac{(n_{k_0+Q}-n_{k_0})P(\omega,2J\sin(Q))}{16J|\cos(2k_0+Q)\sin(Q)|} \nonumber \\
&\quad -\frac{1}{2}(1-\cos(2k_++Q))\frac{(n_{k_++Q}+n_{k_+}-1)}{16J|\sin(2k_++Q)\cos(Q)|} \nonumber \\
&\qquad \times P(\omega+\Delta J,2J \cos(Q)) \nonumber \\
&\quad +\frac{1}{2}(1-\cos(2k_-+Q))\frac{(n_{k_-+Q}+n_{k_-}-1)}{16J|\sin(2k_-+Q)\cos(Q)|} \nonumber \\
& \qquad \times P(\omega-\Delta J,2J \cos(Q)),
\end{align}
with
\begin{align}
2k_0+Q & = \arcsin\left( \frac{\omega}{2J\sin(Q)} \right),\\
2k_++Q & = \arccos\left[-\frac{1}{2\cos(Q)}\left(\frac{\omega}{J}+\Delta \right)\right],\\
2k_-+Q & = \arccos\left[\frac{1}{2\cos(Q)}\left(\frac{\omega}{J}-\Delta \right)\right].
\end{align}
Using the expansion of $\epsilon_k$ one can also show that at next to leading order
\begin{widetext}
\begin{align}
n_{k_0+Q}-n_{k_0}
& = \frac{\sinh(\frac{1}{2} \beta \omega) P(\omega,2J\sin(Q))}{\cosh(\frac{1}{2} \beta \omega)+\cosh(\frac{\beta}{2} J \Delta - \frac{\beta}{2} \cot(Q)((2J \sin(Q))^2-\omega ^2 )^{1/2})}, \nonumber \\ \nn
n_{k_\pm+Q}+n_{k_\pm}-1
& = -\frac{\cosh(\frac{1}{2} \beta \omega) P(\omega \pm J\Delta,2J\cos(Q))}{\cosh(\frac{1}{2} \beta \omega)+\cosh(\frac{\beta}{2}\tan(Q)((2J \cos(Q))^2-(\omega \pm J \Delta )^2 )^{1/2})}.
\end{align}
Inserting these relations into the full expression yields
\begin{align}
-\mathrm{Im}&\chi^{xx}_{R,0} (Q,\omega) \nonumber \\
\approx & \frac{1}{8}\left[\frac{1}{((2J\sin(Q))^2-\omega^2)^{1/2}}-\frac{1}{2J\sin(Q)} \right] \frac{\sinh(\frac{1}{2} \beta \omega) P(\omega,2J\sin(Q))}{\cosh(\frac{1}{2} \beta \omega)+\cosh( \frac{\beta}{2} J \Delta - \frac{\beta}{2} \cot(Q)((2J \sin(Q))^2-\omega ^2 )^{1/2})} \nonumber \\
& + \frac{1}{32J\cos(Q)}\left(\frac{2J\cos(Q)+(\omega+J\Delta)}{2J\cos(Q)-(\omega+J\Delta)}\right)^{\frac{1}{2}}\frac{\cosh(\frac{1}{2} \beta \omega)P(\omega+J\Delta,J\cos(Q))}{\cosh(\frac{1}{2} \beta \omega)+\cosh(\frac{\beta}{2}\tan(Q)((2J \cos(Q))^2-(\omega + J \Delta )^2 )^{1/2})} \nonumber \\
& - \frac{1}{32J\cos(Q)}\left(\frac{2J\cos(Q)-(\omega-J\Delta)}{2J\cos(Q)+(\omega-J\Delta)}\right)^{\frac{1}{2}}\frac{\cosh(\frac{1}{2} \beta \omega)P(\omega-J\Delta,2J\cos(Q))}{\cosh(\frac{1}{2} \beta \omega)+\cosh(\frac{\beta}{2}\tan(Q)((2J \cos(Q))^2-(\omega - J \Delta )^2 )^{1/2})}.
\end{align}
\label{treelevelSxx}
\end{widetext}
This anaysis shows that, at zeroth order, the response diverges as an inverse square root both for $\omega=\pm(\Delta J- 2J\cos(Q))$ and $\omega=\pm2J\sin(Q)$. These divergences are associated with the gapped two--spinon response and the thermally activated Villain mode respectively.
\subsection{First Order Perturbation Theory}
At first order two classes of contributions appear,
which may be distinguished by considering their diagrammatic
representation. As we have already seen, the three zeroth order
diagrams take the form of bubbles. The first class of diagrams in
first order consists of two bubbles connected by an interaction
vertex. The second class of diagrams consists of a single bubble with
a self--energy insertion. We detail these contributions in Appendix
\ref{apponevertex}. An important feature is that several of the first
order contributions have stronger singularities as functions of the
external frequency and momentum than the zeroth order results. This
indicates that it is necessary for the expansion to be resummed before
useful physical results can be extracted.  
\section{Bubble Summation}
\label{secxxzresum}
We have shown that at zeroth order the susceptibility diverges for
certain $\omega$ and $Q$ and that this is matched by stronger
divergences in some of the first order terms (see Appendix
\ref{apponevertex}). Moreover, it is clear from the first order
calculation that higher orders in perturbation theory will exhibit
stronger and stronger divergences. In order to get physically meaningful
results we therefore should sum the most divergent classes of diagrams.
In the case at hand, the complicated momenta dependence of the vertices
and the self--consistent Bogoliubov transformation makes the
determination of the most divergent contributions an impossible
task. Instead we resum just the connected bubble diagrams and
justify our choice by comparing results with the exact $T=0$
calculation. We explain how to incorporate self--energy corrections in
Sec. \ref{secxxzsp}.
\subsection{RPA--Like Scheme}
\label{secxxzrpa}
The RPA scheme consists of carrying out  bubble sums of the type shown in
Fig.~\ref{bubblesum} 
\begin{center}
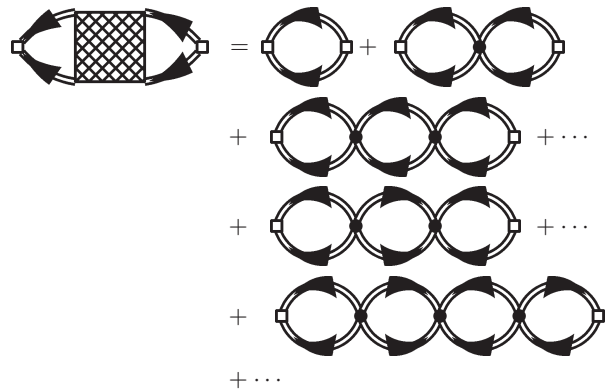
\begin{figure}[ht]
\begin{align*}
\parbox{30mm}{\begin{fmfgraph*}(70,30)
      \fmfleft{v1} \fmfright{v2}
      \fmfpolyn{hatched}{b}{4}
      \fmf{heavy,right=0.2,tension=6}{b1,v1}
      \fmf{heavy,left=0.2,tension=6}{b2,v1}
      \fmf{heavy,left=0.2,tension=6}{v2,b3}
      \fmf{heavy,right=0.2,tension=6}{v2,b4}
      \fmfv{decor.shape=square,decor.filled=empty,decor.size=2thick}{v1,v2}
  \end{fmfgraph*}}
&=\;\parbox{12mm}{\begin{fmfgraph*}(30,30)
      \fmfleft{v1} \fmfright{v2}
      \fmf{heavy,left=0.8}{v2,v1}
      \fmf{heavy,right=0.8}{v2,v1}
      \fmfv{decor.shape=square,decor.filled=empty,decor.size=2thick}{v1,v2}
  \end{fmfgraph*}}
+\;\parbox{24mm}{\begin{fmfgraph*}(60,30)
      \fmfleft{v1} \fmfright{v3}
      \fmf{heavy,left=0.8}{v2,v1}
      \fmf{heavy,right=0.8}{v2,v1}
      \fmf{heavy,left=0.8}{v3,v2}
      \fmf{heavy,right=0.8}{v3,v2}
      \fmfdot{v2}
      \fmfv{decor.shape=square,decor.filled=empty,decor.size=2thick}{v1,v3}
  \end{fmfgraph*}}\\
&+\;\parbox{36mm}{\begin{fmfgraph*}(90,30)
      \fmfleft{v1} \fmfright{v4}
      \fmf{heavy,left=0.8}{v2,v1}
      \fmf{heavy,right=0.8}{v2,v1}
      \fmf{heavy,left=0.8}{v3,v2}
      \fmf{heavy,right=0.8}{v3,v2}
      \fmf{heavy,left=0.8}{v4,v3}
      \fmf{heavy,right=0.8}{v4,v3}
      \fmfdot{v2}
      \fmfdot{v3}
      \fmfv{decor.shape=square,decor.filled=empty,decor.size=2thick}{v1,v4}
  \end{fmfgraph*}}
 +\cdots \nn
&+\;\parbox{36mm}{\begin{fmfgraph*}(90,30)
      \fmfleft{v1} \fmfright{v4}
      \fmf{heavy,left=0.8}{v2,v1}
      \fmf{heavy,right=0.8}{v2,v1}
      \fmf{heavy,left=0.8}{v2,v3}
      \fmf{heavy,right=0.8}{v2,v3}
      \fmf{heavy,left=0.8}{v4,v3}
      \fmf{heavy,right=0.8}{v4,v3}
      \fmfdot{v2}
      \fmfdot{v3}
      \fmfv{decor.shape=square,decor.filled=empty,decor.size=2thick}{v1,v4}
  \end{fmfgraph*}}
+\cdots \nn
&+\;\parbox{48mm}{\begin{fmfgraph*}(120,30)
      \fmfleft{v1} \fmfright{v5}
      \fmf{heavy,left=0.8}{v2,v1}
      \fmf{heavy,right=0.8}{v2,v1}
      \fmf{heavy,left=0.8}{v2,v3}
      \fmf{heavy,right=0.8}{v2,v3}
      \fmf{heavy,left=0.8}{v4,v3}
      \fmf{heavy,right=0.8}{v4,v3}
      \fmf{heavy,left=0.8}{v5,v4}
      \fmf{heavy,left=0.8}{v4,v5}
      \fmfdot{v2}
      \fmfdot{v3}
      \fmfdot{v4}
      \fmfv{decor.shape=square,decor.filled=empty,decor.size=2thick}{v1,v5}
  \end{fmfgraph*}}\nn
& + \cdots
\end{align*}
\caption{Bubble summation for one contribution to the dynamical
  susceptibility matrix 
$\chi^{xx}(i\omega_n,Q)$. The thick lines indicate that the single particle propagators may be resummed to include self--energy corrections.}
\label{bubblesum}
\end{figure}
\end{center}
To carry out these summations is non--trivial as the lines in the
internal bubbles can take any orientation and the momentum dependence
of the vertices is very complicated. In order to proceed we organize
the interaction vertices into a $3\times 3$ matrix
\begin{align}
V_{11}(Q|k,q)&={V}_2(k+Q,-k,q,-q-Q), \nn 
V_{12}(Q|k,q)&= -3{V}_1(q+Q,-q,k,-k-Q), \nn 
V_{13}(Q|k,q)&= 6 {V}_0(k+Q,-k,q,-q-Q), \nn 
V_{21}(Q|k,q)&= 3{V}_1(k+Q,-k,q,-q-Q), \nn 
V_{22}(Q|k,q)&= -4{V}_2(k+Q,q,-k,-q-Q), \nn 
V_{23}(Q|k,q)&= -3{V}_1(k,q+Q,-q,-k-Q), \nn 
V_{31}(Q|k,q)&=6 {V}_0(q+Q,-q,k,-k-Q), \nn 
V_{32}(Q|k,q)&= 3{V}_1(q,k+Q,-k,-q-Q), \nn 
V_{33}(Q|k,q)&= {V}_2(q,-q-Q,k+Q,-k).
\end{align}
As is shown in Appendix \ref{secrpaproof}, in the thermodynamic limit
the RPA--like bubble summation without taking into account self--energy
corrections results in an integral equation of the form
\begin{widetext}
\be
{\Pi}_{\alpha\beta}^{\rm RPA}(i\omega_n,Q|k,k')={\Pi}_{\alpha\beta}^0(i\omega_n,Q|k,k')
+\int\frac{dq}{2\pi}K_{\alpha\gamma}(i\omega_n,Q|k,q)
{\Pi}_{\gamma\beta}^{\rm RPA}(i\omega_n,Q|q,k')\ .
\label{inteq1}
\ee
\end{widetext}
where the kernel $K$ is defined as the infinite volume limit of
\be
K_{\alpha\beta}(i\omega_n,Q|k,q)=\sum_{k'}\Pi^0_{\alpha\gamma}(i\omega_n,Q|k,k')
V_{\gamma\beta}(Q|k',q). 
\label{kernel}
\ee
Defining a convolution $*$ by
\begin{widetext}
\begin{align}
(X*Y)_{\alpha\beta}(i\omega_n,Q,k,k')\equiv
\int_{-\pi}^\pi\frac{dq}{2\pi} X_{\alpha\gamma}(i\omega_n,Q|k,q)Y_{\gamma\beta}(i\omega_n,Q|q,k'),
\end{align}
\end{widetext}
this can be rewritten as 
\begin{align}
\big(\boldsymbol{I}-\boldsymbol{K}\big)*\boldsymbol{\Pi}^{\rm
  RPA}=\boldsymbol{\Pi}^0, 
\label{inteq2}
\end{align}
The integral equation (\ref{inteq2}) is then readily solved
\begin{align}
\boldsymbol{\Pi}^{\rm  RPA}=
\big(\boldsymbol{I}-\boldsymbol{K}\big)^{-1}*\boldsymbol{\Pi}^0.
\end{align}
After analytic continuation to real frequencies
{$i\omega_n\rightarrow \omega+i\eta$} the quantity of
interest is calculated as 
\begin{align}
\int_{-\pi}^\pi\frac{dk\ dq}{(2\pi)^2} L_\alpha(k)
{\Pi}^{\rm RPA}_{\alpha\beta}(\omega+i\eta,Q|k,q) L^\dagger_\beta(q).
\label{eqnresum}
\end{align}
In practice the solution of the integral equation discussed above is
reduced to a simple matrix inversion problem. We discretize all
momentum integrals in terms of sums over $N=400$ points.
We find that this value is large enough to make discretization effects
negligible. 
We set $J=1$ and the regulator $\eta=10^{-3}$. The discretized
representation of $\boldsymbol{I}-\boldsymbol{\Pi}^0*\boldsymbol{V}$
(which is a matrix both in momentum space as well as in the $3\times
3$ matrix space labelled by greek indices) may be found using standard
linear algebra routines. The dynamical structure factor is then
evaluated at  $24000$ points in frequency space. Using a finite system
size results in a finite number of poles in the susceptibility
(\ref{eqnxxzsuscept}). In turn, because $\eta$ is finite, the
calculated structure factor will be composed of a number $\sim N$ of
Lorentzian peaks of width $\eta$. Finally we convolve this result with a
suitable Gaussian. 
\subsection{Comparison with Exact Results for $T=0$}
\begin{figure}[tb]
\begin{center}
\epsfxsize=0.45\textwidth
\epsfbox{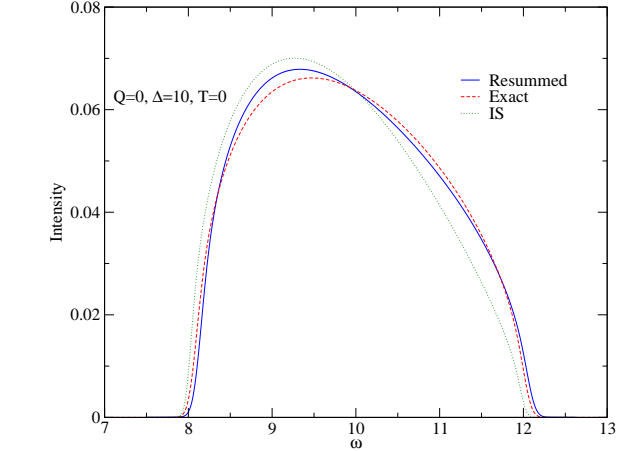}
\caption{(Color Online) The dynamical structure factor as found by resummation, the exact result\cite{bg,isaac} and the calculation of Ishimura and Shiba\cite{is} (IS) at $T=0$, $Q=0$ and $\Delta=10$. In all cases the curves are convolved with a Gaussian in frequency space of full width half maximum 0.12.} 
\label{figxxzlike4like0}
\end{center}
\end{figure}

\begin{figure}[tbh]
\begin{center}
\epsfxsize=0.45\textwidth
\epsfbox{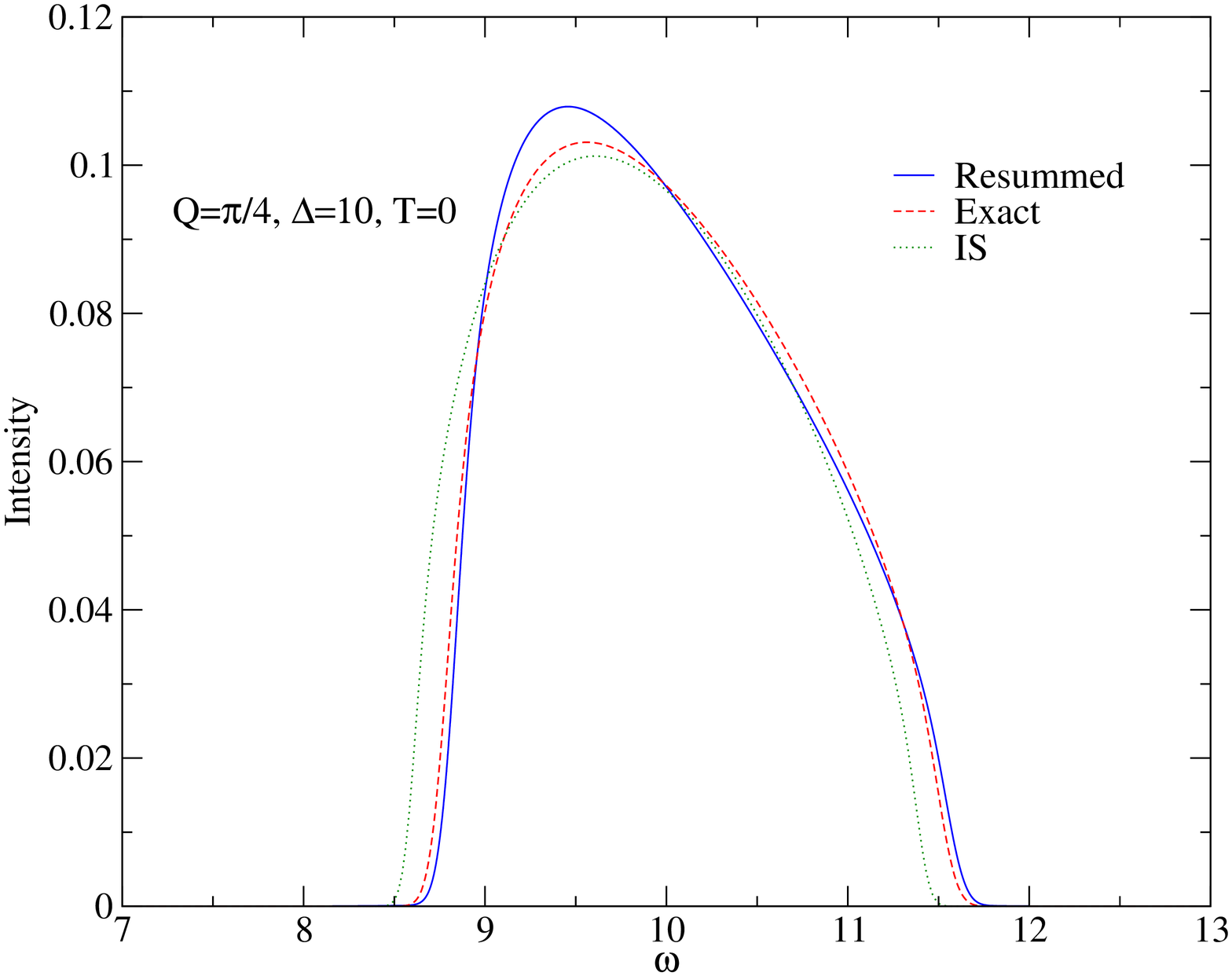}
\caption{(Color Online) The dynamical structure factor as found by resummation, the exact result\cite{bg,isaac} and the calculation of Ishimura and Shiba\cite{is} (IS) at $T=0$, $Q=\pi/4$ and $\Delta=10$. In all cases the curves are convolved with a Gaussian in frequency space of full width half maximum 0.12.} 
\label{figxxzlike4likepi_4}
\end{center}
\end{figure}

\begin{figure}[tbh]
\begin{center}
\epsfxsize=0.45\textwidth
\epsfbox{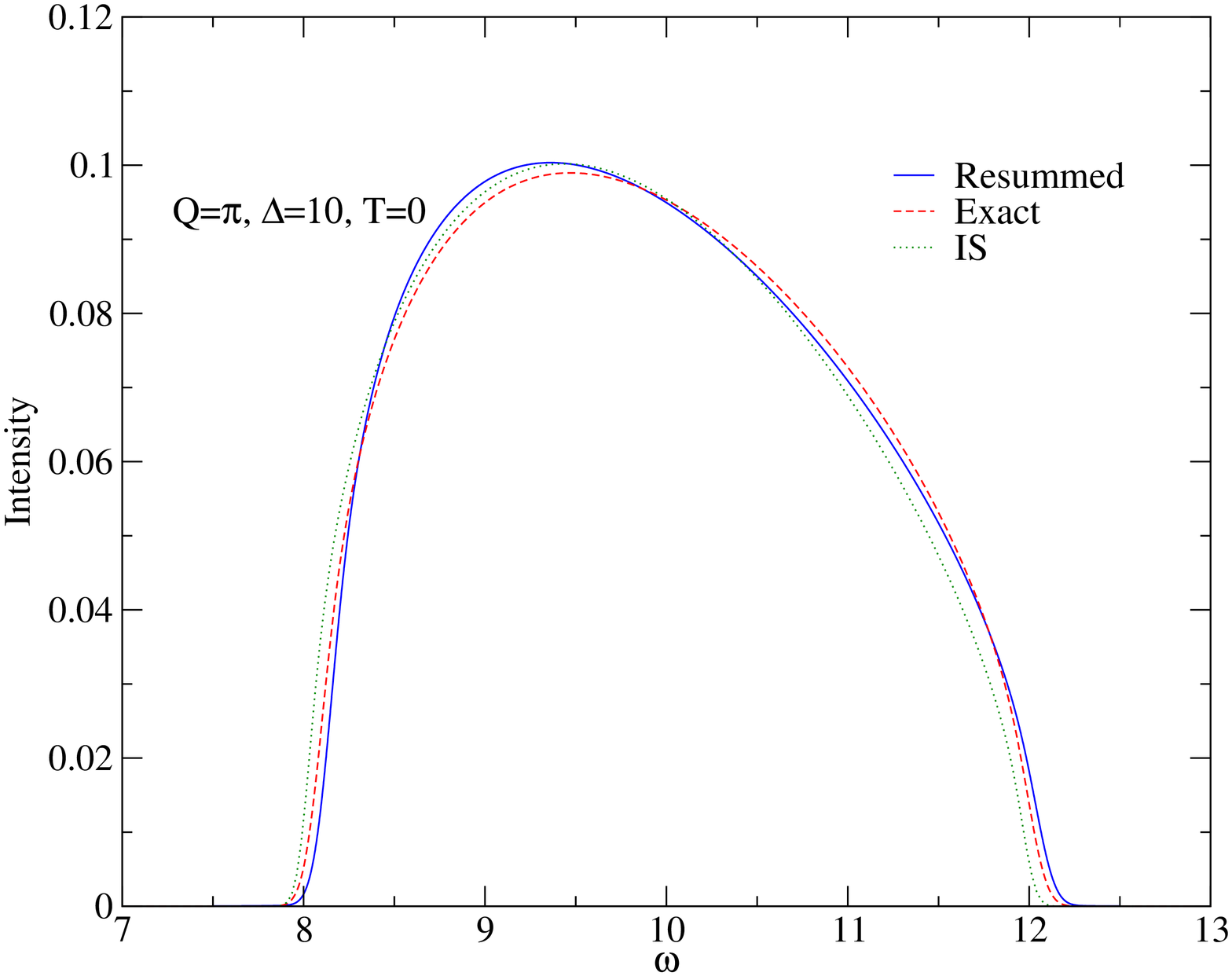}
\caption{(Color Online) The dynamical structure factor as found by resummation, the exact result\cite{bg,isaac} and the calculation of Ishimura and Shiba\cite{is} (IS) at $T=0$, $Q=\pi$ and $\Delta=10$. In all cases the curves are convolved with a Gaussian in frequency space of full width half maximum 0.12.} 
\label{figxxzlike4likepi}
\end{center}
\end{figure}

\begin{figure}[ht]
\begin{center}
\epsfxsize=0.45\textwidth
\epsfbox{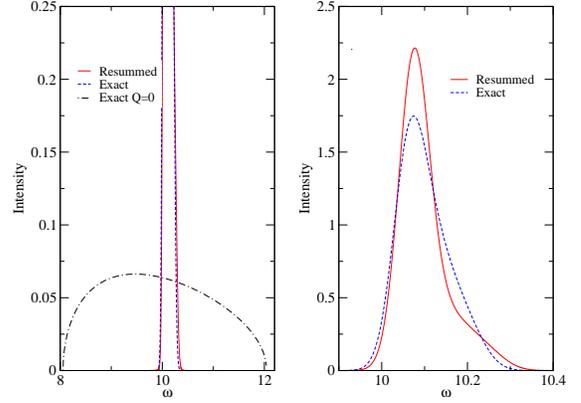}
\caption{(Color Online) A comparison of our calculation and the exact result\cite{bg,isaac} for $Q=\pi/2$ at $T=0$ and $\Delta=10$. Left panel: $\pi/2$ curves convolved with a Gaussian (width 0.08) plotted with the exact result at $Q=0$, in order to demonstrate the scale. Right panel: The same $\pi/2$ curves but plotted at a different scale. At $Q=\pi/2$ the result of Ishimura and Shiba is undefined as the denominator goes to zero.\cite{is}} 
\label{figxxzcomppi_2}
\end{center}
\end{figure}

Given the uncontrolled nature of our bubble summation it is essential
to compare it to exact results at zero temperature \cite{bg,isaac} in
order to assess its quality. In Figs. \ref{figxxzlike4like0},
\ref{figxxzlike4likepi_4}, \ref{figxxzlike4likepi} and \ref{figxxzcomppi_2} we plot our results against the exact results for the dynamical structure factor for $\Delta=10$,
$T=0$ and several momenta. We also include for comparison an earlier result for the DSF due to Ishimura and Shiba.\cite{is}
We see that for $\Delta=10$, $T=0$ and at most wave vectors the
resummation is a highly accurate approximation. This suggests that the
diagrams we have selected account for most of the spectral weight at low
temperature. It is known from the exact result\cite{bg,isaac} that the
gapped response diverges along its lower energy threshold, in a region
of momentum centred about $\pi/2$ (the size of which increases as
$\Delta \to 1$). Correspondingly the resummation is less accurate near
$\pi/2$, although there is still qualitative agreement
(Fig. \ref{figxxzcomppi_2}). For larger values of $\Delta$ our
approximation becomes even better.
\subsection{Analytical Resummation for $T=0$}
It is possible and instructive to evaluate the resummation exactly in
the limit $T=0$, $\Delta \to \infty$. At zeroth order, positive
frequencies and $T=0$, the only non--zero diagram is the
particle--particle propagation bubble. This diagram leads to a
response in the region $\omega \sim \Delta J$. 
As $T \rightarrow 0$ the thermal occupation factors vanish and by
expanding in $1/ \Delta$ we find 
\be
\theta_k = {\cal O}(\Delta^{-1}),
\ee
so that we can neglect the Bogoliubov phases.
The diagram then reduces to 
\be
\int \frac{dk}{4\pi}\ \frac{\sin^2(k+Q/2)}{i
  \omega_n-\epsilon_k-\epsilon_{k+Q}}=A(i\omega_n,Q).
\ee
Using the approximation
\[\epsilon_k+\epsilon_{k+Q}\approx\Delta J +2J \cos(Q) \cos(2k+Q)\]
we have 
\begin{align*}
A(i\omega_n,Q) = & \int_{-\pi}^{\pi} \frac{dk}{4\pi}\ \frac{ \sin^2(k+Q/2)}{i \omega_n-\epsilon_k-\epsilon_{k+Q}} \\
= & \frac{1}{4J \cos(Q)} \int_{0}^{\pi} \frac{dk}{2\pi} \frac{1-\cos(2k+Q)}{\tilde{\omega}-\cos(2k+Q)}
\end{align*}
and $ \tilde{\omega}  =  \frac{i \omega_n-\Delta J}{2J \cos(Q)}$. The
remaining integral can be taken by standard contour integration methods
and analytic continuation to real frequencies is then straightforward.
With the approximations made here the vertex $V_2$ takes a
particularly simple form and resummation amounts to the geometric
series 
\begin{align}
-\mathrm{Im}[\chi^{xx}(\omega,Q)] = &-\mathrm{Im}\big( A(\omega,Q)+4J \cos(Q) A(\omega,Q)^2\nn
&+(4J)^2 \cos^2(Q) A(\omega,Q)^3 + \hdots\big) \nn
= &-\mathrm{Im}( \frac{A(\omega,Q)}{1-4J \cos(Q) A(\omega,Q)}).
\end{align}
Finally the $T=0$ result, to lowest order in $\Delta$ is
\begin{widetext}
\begin{align}
\label{eqnxxzT0exact}
-\mathrm{Im}[\chi^{xx}(\omega,Q)]= & \begin{cases}\frac{\sqrt{(2J\cos(Q))^2-(\omega-\Delta J)^2}}{8 J^2 \cos^2(Q)}&  |\omega-\Delta J| \le |2J\cos(Q)|\\
0 & \text{otherwise.}
\end{cases}
\end{align}
\end{widetext}
Ishimura and Shiba calculated the DSF at $T=0$ using a method based on
perturbation theory combined with a Lehmann representation.\cite{is}
The expression obtained here (\ref{eqnxxzT0exact}) agrees to lowest
order with their result.  
\subsection{One--Loop Self--Energy Corrections}
\label{secxxzsp}
As they stand, the first order tadpole contributions,
Eqs. (\ref{tadpole1}--\ref{tadpole7}), cannot be incorporated into the
RPA. This is because they contain divergences which are not resummed
according to the scheme above. Instead we must first calculate a new
single particle Green's function, using a Dyson equation to resum the
tadpole self--energy corrections. This Green's function is then used
to calculate a new bubble diagram.  

Including the tadpole corrections generates anomalous propagators at
higher orders. These are most efficiently taken into account by
formulating the propagator as a matrix, 
\begin{align}
&\boldsymbol{g}(i\omega_n,k)=-\int_0^\beta d\tau e^{i\omega_n \tau}
\boldsymbol{g}(\tau,k)\ ,\nn
&\boldsymbol{g}(\tau,k)= \bigg\langle T_\tau\begin{pmatrix}
\alpha_k(\tau)\alpha_k^\dagger(0) & \alpha_k(\tau)\alpha_{-k}(0) \\
\alpha_{-k}^\dagger(\tau)\alpha_k^\dagger(0) & \alpha_{-k}^\dagger(\tau)\alpha_{-k}(0)
\end{pmatrix} U(\beta,0)\bigg\rangle
\end{align}
At zeroth order the result is
\begin{align}
\boldsymbol{g}_0(i\omega_n,k) = \left( \begin{array}{cc}
G_0(i\omega_n,k) & 0 \\
0 & -G_0(-i\omega_n,-k)
\end{array}\right).
\end{align}
The single particle self--energy is defined by the Dyson equation
\begin{align}
\boldsymbol{g}^{-1}(i\omega_n,k)& = \boldsymbol{g}_0^{-1}(i\omega_n,k)
-\boldsymbol{\Sigma}(i\omega_n,k). 
\end{align}
If only `tadpole' type diagrams are included the self--energy is, to
first order in perturbation theory, frequency independent
\begin{align}
\boldsymbol{\Sigma} = \sum_p \frac{n_p}{N} \left( \begin{array}{cc}
4{V}_2(k,p,-p,-k) & -3{V}_1(p,-p,k,-k) \\
3{V}_1(p,-p,k,-k) & -4{V}_2(k,p,-p,-k)
\end{array}\right).
\end{align}
The elements of the matrix are $\mathcal{O}(n_k \Delta^0)$ and hence their effects will be most pronounced when the gap is small or the temperature is large.

Using the relations
\begin{align}
g_0^{11}(i\omega_n,k)& = \frac{1}{i\omega_n-\epsilon_k}, \\
g_0^{22}(i\omega_n,k)& = \frac{1}{i\omega_n+\epsilon_k}, \\
\Sigma^{21} & = \big(\Sigma^{12}\big)^\star = -\Sigma^{12}, \\
\Sigma^{11} & = -\Sigma^{22},
\label{1loopprops}
\end{align}
one finds that
\begin{multline}
\boldsymbol{g}(i\omega_n,k)
 = \frac{1}{(i\omega_n)^2-(\epsilon_k
 +\Sigma^{11})^2-\lvert \Sigma^{12}\rvert^2}\nn
\times\left( \begin{array}{cc}
i\omega_n+\epsilon_k +\Sigma^{11}& -\Sigma^{12} \\
\Sigma^{12} & i\omega_n-\epsilon_k-\Sigma^{11}
\end{array}\right).
\end{multline}
We rewrite this as
\begin{align}
g^{11}(i\omega_n,k) & = \Big[
  \frac{Z_k^-}{i\omega_n-E_k}+\frac{Z^+_k}{i\omega_n+E_k}  \Big], \\ 
g^{22}(i\omega_n,k) & = \Big[
  \frac{Z^-_k}{i\omega_n+E_k}+\frac{Z^+_k}{i\omega_n-E_k}  \Big], \\ 
g^{12}(i\omega_n,k) & = \lambda_k\Big[
  \frac{1}{i\omega_n+E_k}-\frac{1}{i\omega_n-E_k}  \Big], \\ 
g^{21}(i\omega_n,k) & =-g^{12}(i\omega_n,k),
\end{align}
with the definitions
\begin{align}
E_k & = \sqrt{(\epsilon_k+\Sigma^{11})^2+\lvert \Sigma^{12} \rvert^2}, \label{eqnxxzTdisp}\\\nn
Z^\pm_{k} & =\frac{1}{2}\Big( 1\mp\frac{\epsilon_{k}+\Sigma^{11}}{E_{k}}\Big),\quad
\lambda_k  = \frac{\Sigma^{12}}{2E_k}.
\end{align}
\subsection{Bubble Summation with Self--Energy Corrections}
\label{secRPA2}
\begin{figure}[t]
\begin{center}
\epsfxsize=0.45\textwidth
\epsfbox{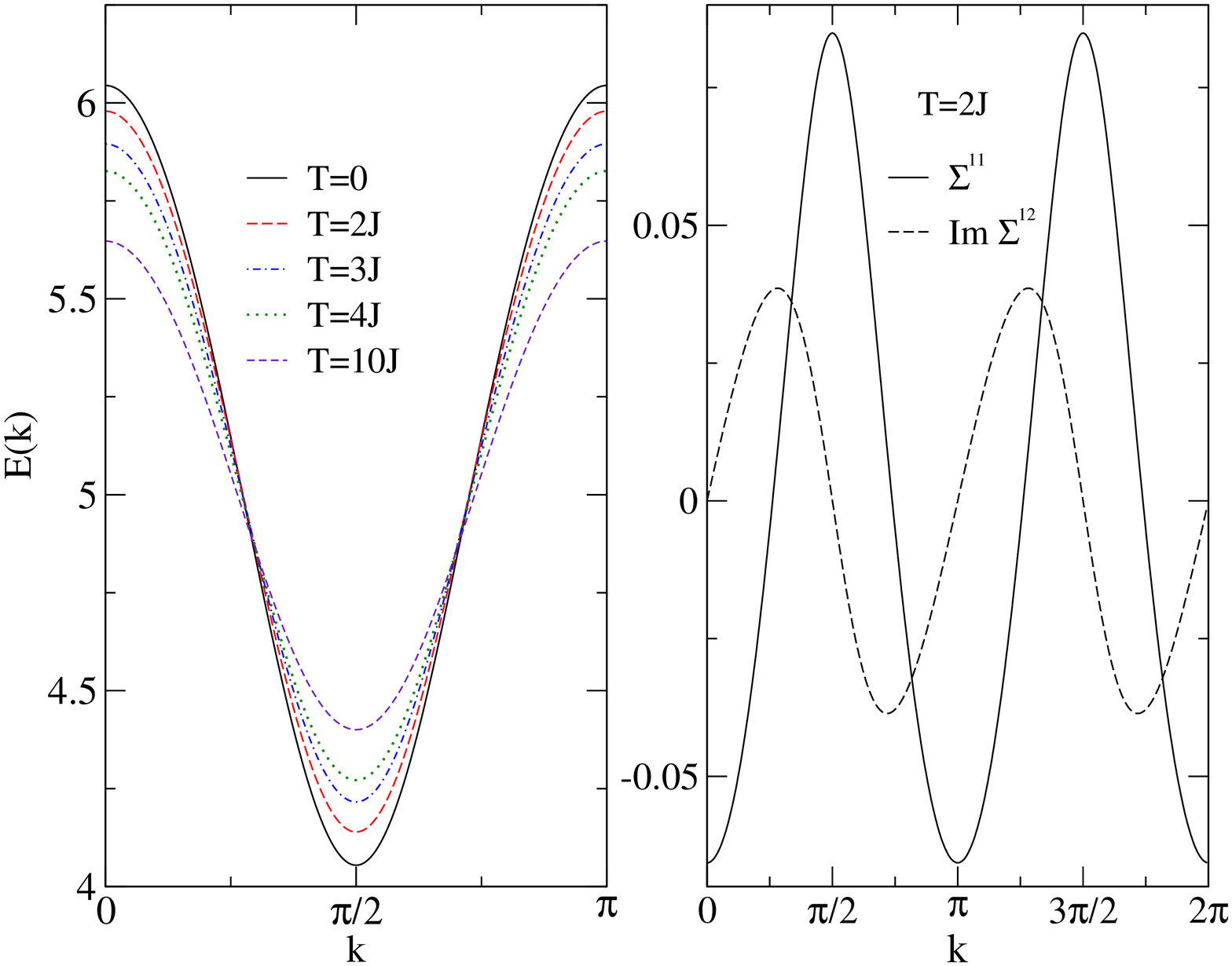}
\caption{(Color Online) The dispersion $E_k$, including tadpole
  self--energy corrections for $\Delta=10$. The left panel shows the
  gradual narrowing of the bandwidth with temperature. At this scale
  the dispersion for $T=J$ is indistinguishable from that for
  $T=0$. The right panel shows the two distinct elements of the self
  energy matrix, $\bm{\Sigma}_{\rm sp}$ at $T=2J$.}  
\label{figxxzdispt}
\end{center}
\end{figure}

The one--loop self--energy corrections to the propagators can be
taken into acount in the bubble summation for the dynamical
susceptibility as follows. We define a $3\times 3$ matrix
$\boldsymbol{\Pi}^S$ by
\bea
&&\Pi^S_{\beta\gamma}(\tau,Q|k,q)  =\nn
&&-
\bigl\langle T_\tau X_{\beta\beta}(\tau,Q|k)\
X^\dagger_{\gamma\gamma}(0,Q|q) U(\beta) 
\bigr\rangle\Big|_{{\rm 1-loop}\ \Sigma} , \label{PiS}
\eea
where only one--loop self--energy corrections are taken into account.
This amounts to calculating the two--point function of $X$ and
$X^\dagger$ using the (anomalous) propagators (\ref{1loopprops}).
The elements of $\boldsymbol{\Pi}^S$ are listed in Appendix
\ref{appelementspis}. We now follow the same steps as
in Appendix \ref{secrpaproof} and show that summing all bubble
diagrams of the form shown in Fig.\ref{bubblesum} gives rise to an
integral equation obtained from (\ref{inteq1}) by the replacement
$\Pi^0_{\alpha\beta}\longrightarrow \Pi^S_{\alpha\beta}$, i.e.
\begin{widetext}
\be
{\Pi}_{\alpha\beta}^{\rm RPA}(i\omega_n,Q|k,k')={\Pi}_{\alpha\beta}^S(i\omega_n,Q|k,k')
+\int\frac{dq}{2\pi}K^S_{\alpha\gamma}(i\omega_n,Q|k,q)
{\Pi}_{\gamma\beta}^{\rm RPA}(i\omega_n,Q|q,k')\ .
\label{inteq1b}
\ee
\end{widetext}
where the kernel $K^S$ is defined as the infinite volume limit of
\be
K^S_{\alpha\beta}(i\omega_n,Q|k,q)=\sum_{k'}\Pi^S_{\alpha\gamma}(i\omega_n,Q|k,k')
V_{\gamma\beta}(Q|k',q). 
\label{kernelb}
\ee
At $T=0$ the one--loop corrections to
$\boldsymbol{\Sigma}$ vanish, so that we recover our previous result.
On the other hand, for $T>0$, $\boldsymbol{\Sigma}$ plays an important
role, altering the dispersion and shifting the thresholds of the
dynamical response (see Fig. \ref{figxxzdispt}). 

One can also calculate the two--loop corrections to the single particle propagator but it is not as simple to incorporate them into the RPA. For reference they are included in Appendix \ref{app2loopsp}.
\section{Results and Discussion}
\label{secxxzresults}
We now turn to a discussion of our results at finite temperatures and
applied fields. For the perturbative expansion to be valid we require
$\Delta \gg 1$. In selecting diagrams to resum we have favoured those
that are most relevant at low temperatures, i.e. those that feature
few thermal occupation factors. The relevant energy scale is the
single particle gap $\sim \Delta J/2$, so we restrict our discussion
to temperatures $T<\Delta J/2$. 
We calculate the transverse dynamical structure factor as described in
Sec. \ref{secxxzrpa}, using the matrix $\boldsymbol{\Pi}_S$ found in
Sec. \ref{secxxzsp} and setting $\Delta=10$, $J=1$.  For $\Delta \gg
1$ the dynamical structure factor (at positive frequency) consists of
two pieces: a gapped continuum response occuring at frequencies
$\omega \sim \Delta$ and a response for $\omega \sim 0$ that is only
seen at finite temperature. 

On general grounds one expects that at finite temperature the very
sharp thresholds seen at $T=0$ should disappear. In our
approach the thresholds are still present although they are obfuscated
by the necessity of convolving the response with a Gaussian. This is a
consequence of the diagrams we have taken into account. We expect this
`thermal broadening' to be a small effect that could be taken into
account by including certain two--loop diagrams. These two--loop
diagrams connect the response to decay channels of higher particle
number. We discuss the issue further in Appendix \ref{app2loopsp}. 

\begin{figure}[tb]
\begin{center}
\epsfxsize=0.45\textwidth
\epsfbox{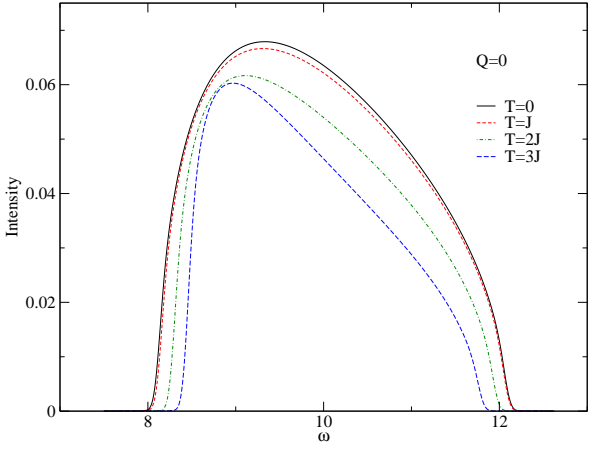}
\caption{(Color Online) The dynamical structure factor at finite temperature for $\omega \sim \Delta$ and $\Delta=10$ at wave vector $Q=0$.} 
\label{figxxzTscana}
\end{center}
\end{figure}
\begin{figure}[tb]
\begin{center}
\epsfxsize=0.45\textwidth
\epsfbox{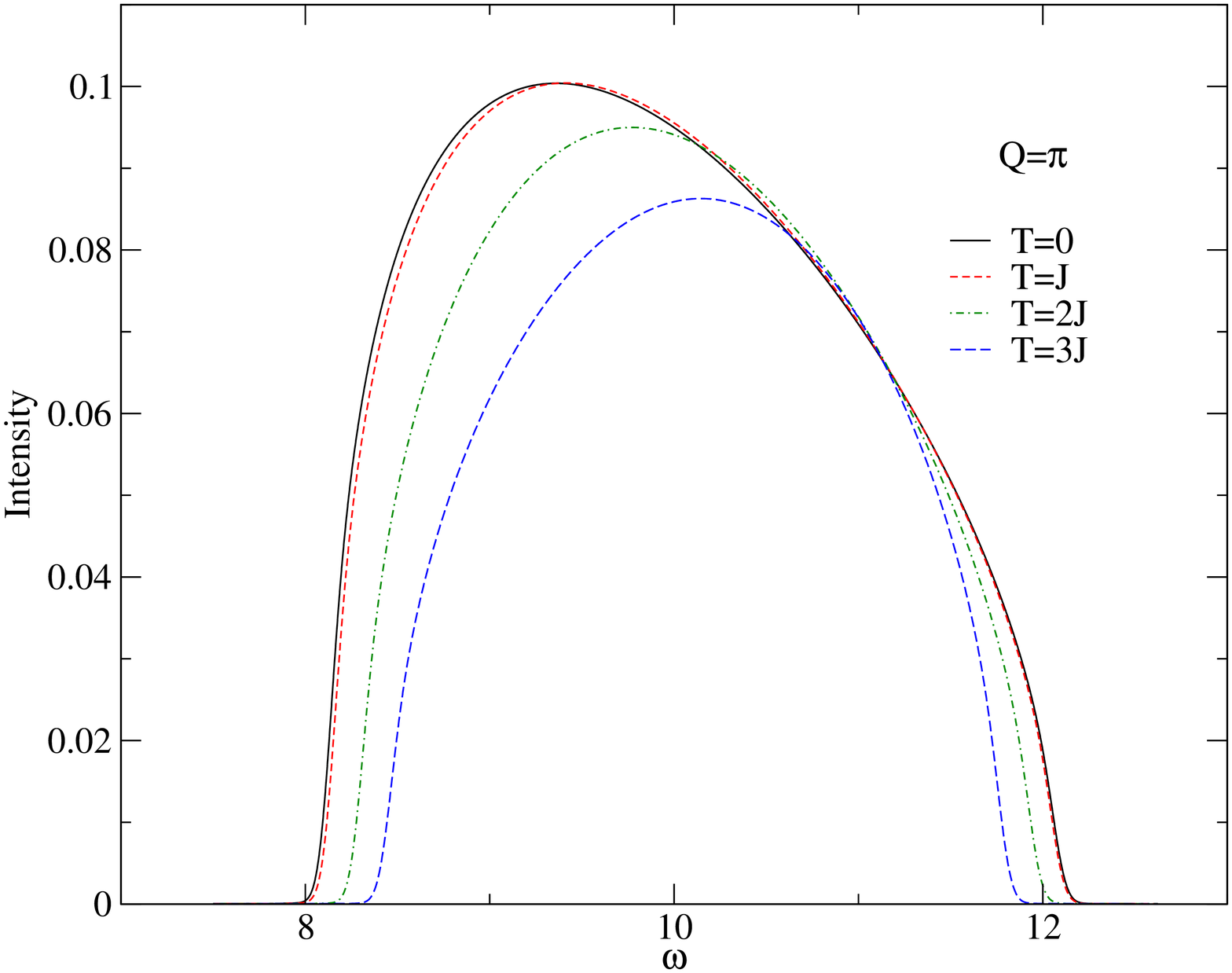}
\caption{(Color Online) The dynamical structure factor at finite temperature for $\omega \sim \Delta$ and $\Delta=10$ at wave vector $Q=\pi$.} 
\label{figxxzTscanb}
\end{center}
\end{figure}
\begin{figure}[tb]
\begin{center}
\epsfxsize=0.45\textwidth
\epsfbox{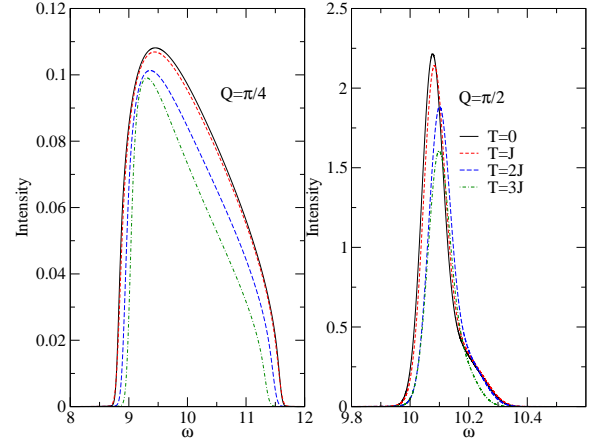}
\caption{(Color Online) The dynamical structure factor at finite temperature for $\omega \sim \Delta$ and $\Delta=10$ at wave vectors $Q=\pi/4$ and $\pi/2$. Note that the scale of the intensity axes differs dramatically between the two plots.} 
\label{figxxzTscan2}
\end{center}
\end{figure} 
\subsection{Gapped Response for $h=0$ }
We first consider the gapped (interband) response at finite
temperatures and for $h=0$. At $T=0$ the transverse DSF is dominated
by a two--spinon scattering continuum that occurs at energies around
twice the single--particle gap, i.e.  $\omega\sim\Delta J$. 
In Figs. \ref{figxxzTscana}, \ref{figxxzTscanb} and \ref{figxxzTscan2}
we show how the DSF in this regime of energies changes at finite
temperature. The most striking feature is a narrowing of the response
with increasing temperature. This is in agreement with inelastic
neutron scattering experiments on ${\rm TlCoCl_3}$\cite{oosawa}.
Another notable feature at $T=0$ is that the response is not
symmetric about $Q=\pi/2$.\cite{isaac} As can be seen in Figs. \ref{figxxzlike4like0}--\ref{figxxzlike4likepi}
our calculation captures this behaviour. Accordingly the response
develops asymmetrically with temperature, and in a non--trivial
way. In particular the maximum of the response for $0\le Q<\pi/2$
moves to lower frequencies as temperature increases, but is shifted to
higher frequencies for $\pi/2<Q\le\pi$. 

For all wave vectors the total spectral weight in the gapped region
decreases as temperature increases. The thresholds of the response
shift as temperature increases due to the thermal dependence of the
single particle dispersion $E_k$, Eq. (\ref{eqnxxzTdisp}). This
depletion of spectral weight is physically sensible because the
excitations are fermionic; as temperature increases more states are
thermally occupied and concomitantly there are fewer states for the
new pair of fermions to fill. 
\subsection{Villain Mode for $h=0$}
\begin{figure}[tb]
\begin{center}
\epsfxsize=0.45\textwidth
\epsfbox{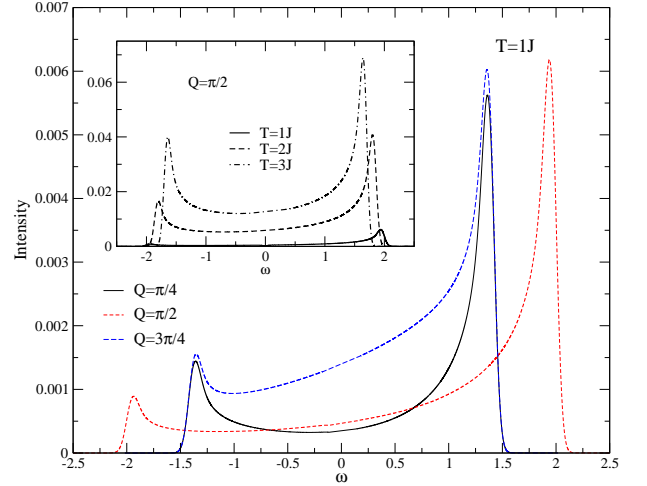}
\caption{(Color Online) The dynamical structure factor for $\omega \sim 0$ and $\Delta=10$. Main plot: the Villain mode for a range of wave vectors. Inset: the development of the Villain mode at $Q=\pi$ with temperature.} 
\label{figxxzvill}
\end{center}
\end{figure}
At temperatures greater than zero there is a thermal population of
spin excitations. Neutron scattering can then lead to processes that
do not change the spinon number of a given microstate, thus giving
rise to an intraband respose at low energies, $\omega \sim
0$. Dynamics of this kind were first described by Villain for the case of
an finite length chain with an odd number of sites.\cite{villain}
As such processes rely on states being thermally occupied, their
contribution to the DSF grows with temperature. The principle feature
of the response is a well defined resonance or `mode' at $\omega=2 J
\sin(Q)$.\cite{villain,nagler,nagler2}
At lowest order in our calculation the intraband response exhibits
a square--root divergence (\ref{treelevelSxx}). Taking interactions
into account by our bubble summation leads to a smoothing of the
divergence, which however occurs in a very small region in 
energy, close to the threshold. This means that the zeroth order
(especially when convolved with an experimental resolution) is an
excellent approximation to the resummed result.

The intraband response is also asymmetric about $Q=\pi/2$ (see Fig. \ref{figxxzvill}). For $0\le Q<\pi/2$ the response between the peaks is suppressed relative to that at $Q=\pi/2$. For 
$\pi/2< Q \le \pi$ it is enhanced.
\subsection{Response in a Transverse Field}
\begin{figure}[tb]
\begin{center}
\epsfxsize=0.45\textwidth
\epsfbox{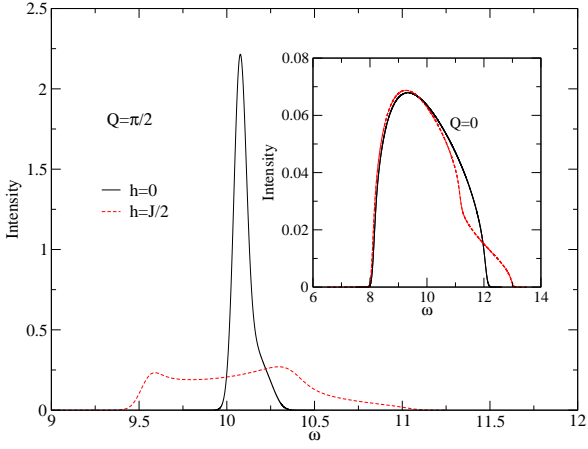}
\caption{(Color Online) Comparison of the $T=0$ dynamical structure factor for $\Delta=10$ with and without an applied transverse field $h=J/2$. Main panel: the divergent response at $\pi/2$ is split into a much broader two peaked structure by the application of the field. Inset: the effect of the field at $Q=0$ is only apparent at the upper threshold.} 
\label{figxxzh1}
\end{center}
\end{figure}
\begin{figure}[tb]
\begin{center}
\epsfxsize=0.45\textwidth
\epsfbox{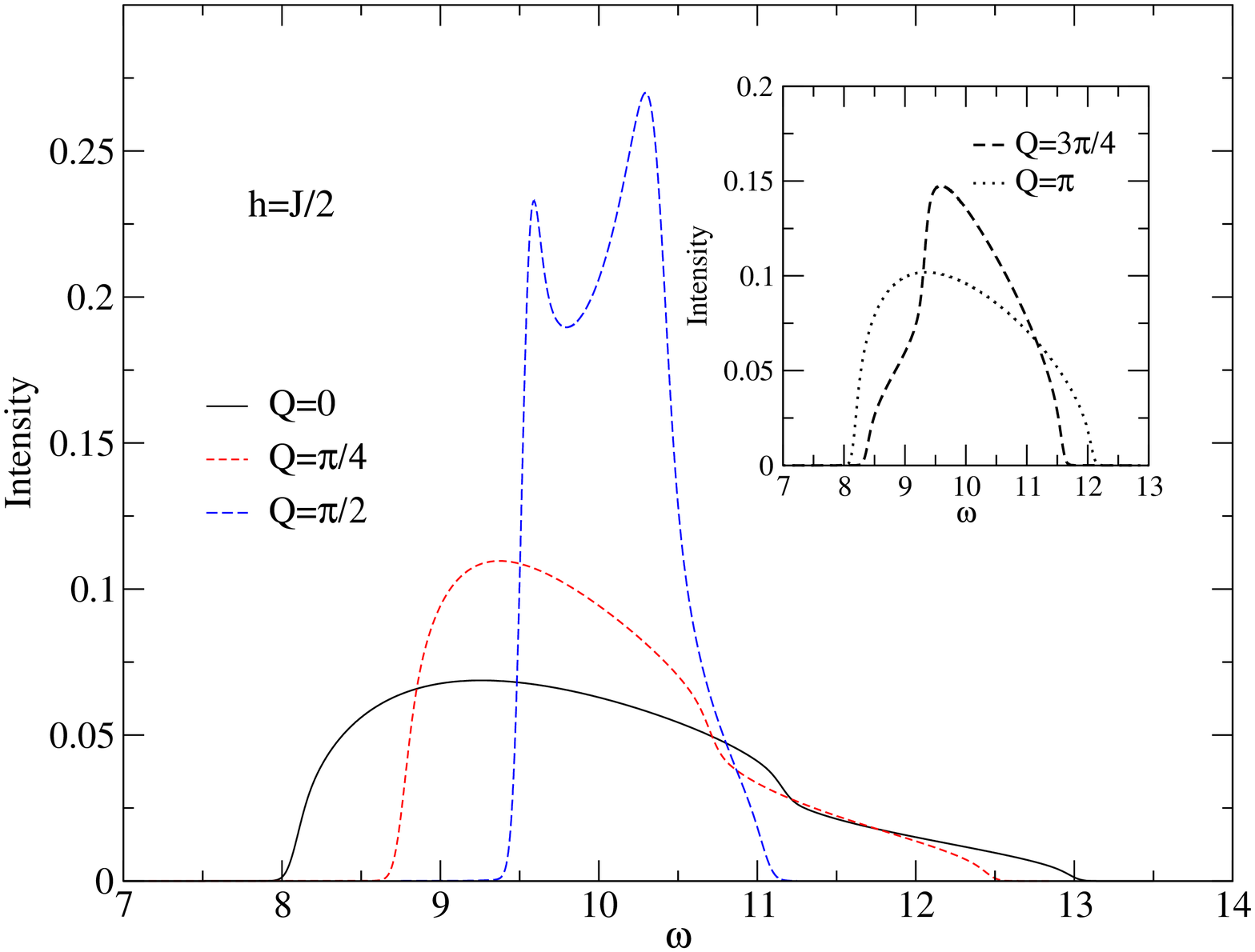}
\caption{(Color Online) Wave vector dependence of the dynamical structure factor for $\Delta=10$ at $h=J/2, T=0$. The main panel shows wave vectors $Q=0,\pi/4$ and $\pi/2$. The inset shows plots for $Q=3\pi/4$ and $\pi$.} 
\label{figxxzh}
\end{center}
\end{figure}
\begin{figure}[tb]
\begin{center}
\epsfxsize=0.45\textwidth
\epsfbox{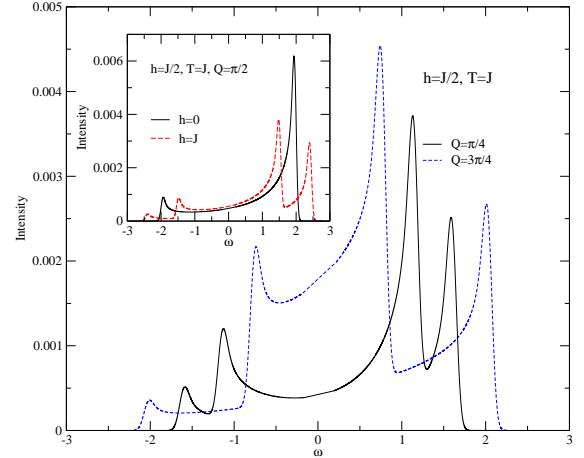}
\caption{(Color Online) Wave vector dependence of the Villain mode for $\Delta=10$ at $h=J/2,T=J$. The main panel shows wave vectors $Q=\pi/4,\pi/2$ and $3\pi/4$. The inset shows the response at $Q=\pi/2$ for $h=0$ and $h=J/2$.} 
\label{figxxzvillh}
\end{center}
\end{figure}
The transverse field $h$ only enters the quadratic part of the
Hamiltonian (\ref{eqnxxzH2}), hence its influence on the scattering
response is through the single particle dispersion and not the
interaction vertices. The first property to take into consideration is
that $h$ will have an effect on the excitation gap. As the magnitude
of the field $h$ approaches $\Delta J$ the gap collapses and the
perturbative expansion is inapplicable. Hence we consider a field
small compared to $\Delta J$. For fields $h \gg \Delta J$ the gap
opens again with the chain ferromagnetically ordered. 
A second important feature to note is that a non--zero $h$ causes the
period of $E_k$ to double (see Fig. \ref{figxxztheta}): the maximum at
$k=(2n+1)\pi$ is reduced relative to those at $k=2n\pi$. This leads to the
double peaked structure seen at $Q=\pi/2$ in the gapped response
(Figs. \ref{figxxzh1}, \ref{figxxzh}) and throughout the low energy
scattering (Fig. \ref{figxxzvillh}). This splitting of the Villain
mode peaks was observed by Braun \emph{et al}.\cite{braun} 

The asymmetry of the response about $Q=\pi/2$ is further increased by
the transverse field. In particular, though the exact result at
$T=0$\cite{bg,isaac} and $\Delta=10$ shows that the energy
\emph{thresholds} of the gapped response are very nearly symmetric
about $\pi/2$, this is no longer the case in a finite field. Instead
the upper threshold near $Q=0$ is pushed to higher energies but is
relatively unchanged near $Q=\pi$. We also note that the narrowing in
energy of the response near $Q=\pi/2$ is suppressed relative to
$h=0$. 
Asymmetry is also seen in the thresholds of the low energy response (Fig. \ref{figxxzvillh}).
\section{Comparison to Diagonalization of Short Chains at $T>0$}
\label{sec:ED}
As we have seen above, at zero temperature our approach gives 
good agreement with the exact DSF. In order to assess the quality of
our approximate DSF at finite temperatures, we have computed the DSF
by means of numerical diagonaliztion of the Hamiltonian on finite periodic
chains of up to 16 sites.
The dynamical susceptibility (\ref{eqnxxzsuscept}) is expressed by
means of a Lehmann expansion in terms of Hamiltonian eigenstates
$|n\rangle$ of energy $E_n$ as
\bea
\chi^{xx}(\omega,Q)&=&N\sum_{n,m} \frac{e^{-\beta
 E_{n}}-e^{-\beta E_{m}}}{\omega+i\eta+E_{n}-E_{m}}\delta_{Q+p_{n},p_{m}}\nn
&&\quad\times\ 
\bra{n}S^x_0\ket{m}\bra{m}S^x_0\ket{n},
\label{Lehmann}
\eea
where the sums are taken over the eigenbasis of $H$.
For a sufficiently small system the eigenstates can be calculated
numerically. Due to the exponential increase in the dimension of the
Hilbert space with the number of sites the method is restricted to
short chains. We consider systems with $N\leq 16$.
In order to approximate the thermodynamic spectrum from a finite
number of allowed transitions, we take the parameter $\eta$ in
(\ref{Lehmann}) to be sufficiently large, so that the finite sum of
Lorentzians in (\ref{Lehmann}) becomes a smooth function of (the real
part of the) frequency. Clearly this procedure can give a meaningful
approximation of the susceptibility in the thermodynamic limit only if
$\eta$ is very small in comparison to the thermal broadening. Hence
the method is restricted to sufficiently large temperatures.
In order to obtain a measure of the importance of finite--size effects
we calculate the DSF for system sizes $N=8,12,16$. We find that the
finite--size effects depend strongly on which region of energy and the
momentum we consider.

\begin{figure}[tb]
\begin{center}
\epsfxsize=0.45\textwidth
\epsfbox{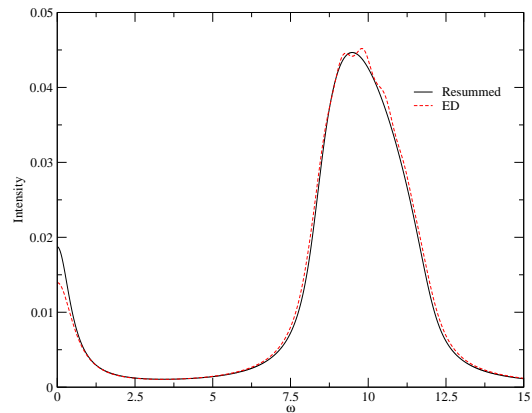}
\caption{(Color Online) Comparison of the dynamical structure factor
  obtained from the perturbative approach to exact diagonalization on a
  16--site chain with $\eta=0.5J$ for $T=2J$ at momentum $Q=0$. In
  order to facilitate a comparison the perturbative result has been
  convolved with a Lorentzian of width $\eta$. }
\label{fig:ed1}
\end{center}
\end{figure}

\begin{figure}[tb]
\begin{center}
\epsfxsize=0.45\textwidth
\epsfbox{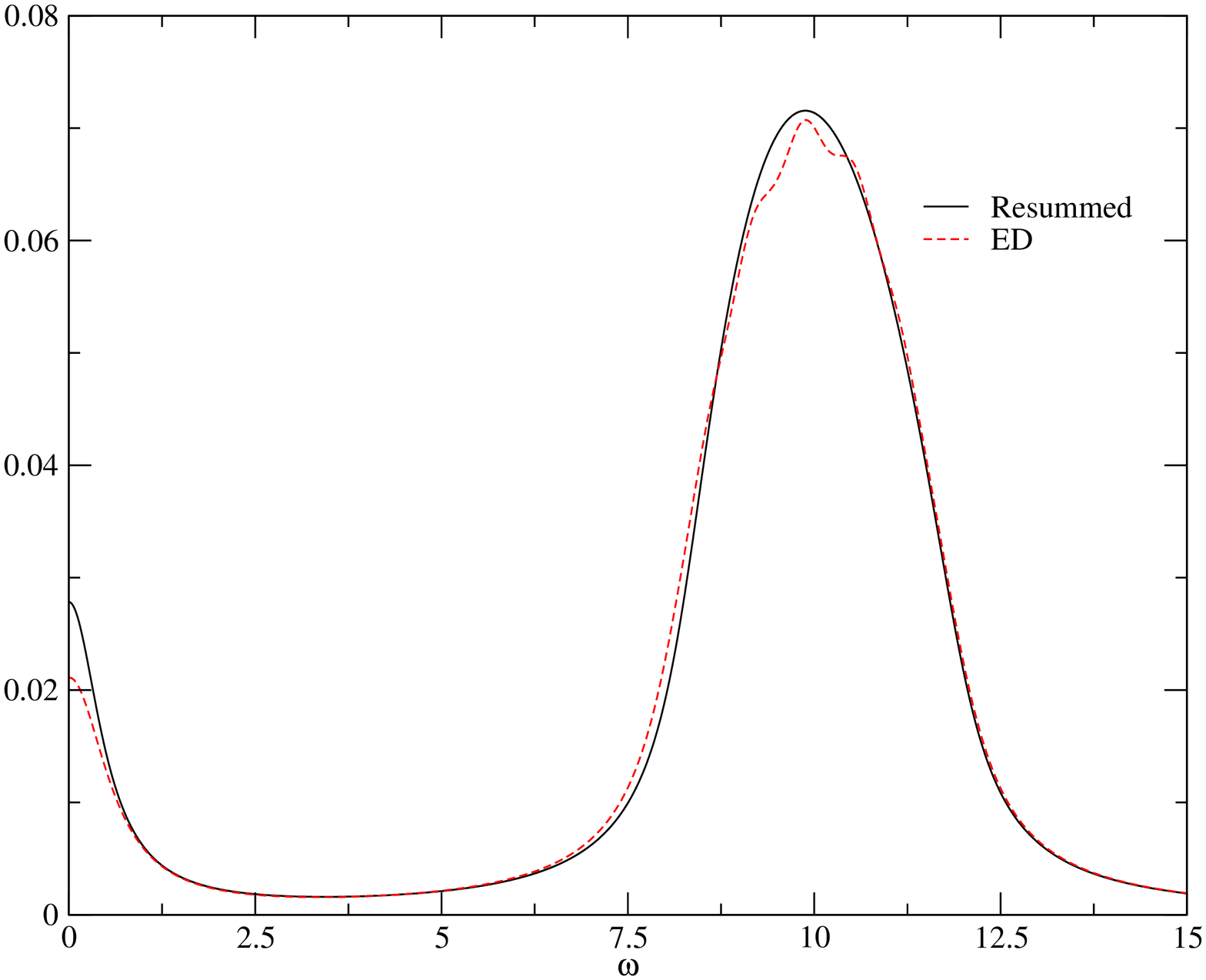}
\caption{(Color Online) Comparison of the dynamical structure factor
  obtained from the perturbative approach to exact diagonalization on a
  16--site chain with $\eta=0.5J$ for $T=2J$ at momentum $Q=\pi$. In
  order to facilitate a comparison the perturbative result has been
  convolved with a Lorentzian of width $\eta$. }
\label{fig:ed2}
\end{center}
\end{figure}

\begin{figure}[tb]
\begin{center}
\epsfxsize=0.45\textwidth
\epsfbox{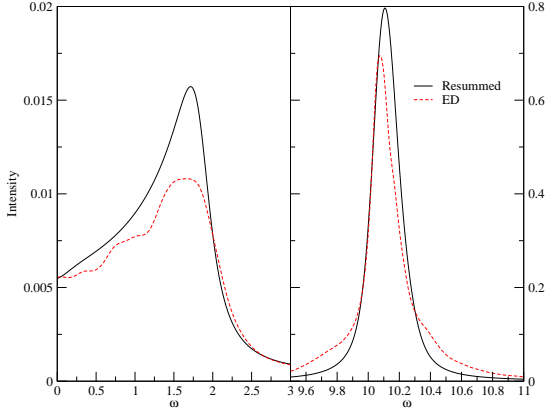}
\caption{(Color Online) Comparison of the dynamical structure factor
  obtained from the perturbative approach to exact diagonalization on a
  16--site chain for $T=2J$ at momentum $Q=\pi/2$. Left panel: low frequency response with $\eta=0.2J$. Right panel: high frequency response with $\eta=0.05J$.
  In order to facilitate a comparison the perturbative results have been
  convolved with a Lorentzian of width $\eta$.}
\label{fig:ed3}
\end{center}
\end{figure}

\begin{figure}[tb]
\begin{center}
\epsfxsize=0.45\textwidth
\epsfbox{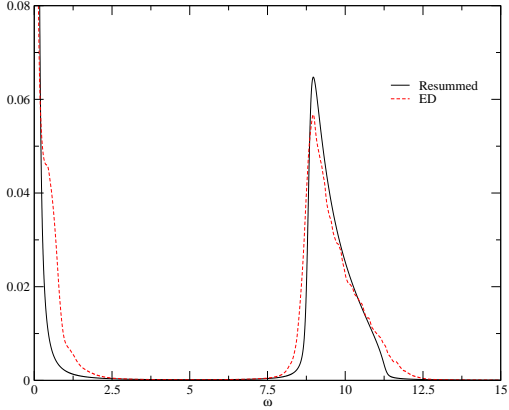}
\caption{(Color Online) Comparison of the dynamical structure factor
  obtained from the perturbative approach to exact diagonalization on a
  16--site chain with $\eta=0.05J$ for $T=10J$ at momentum $Q=0$. In
  order to facilitate a comparison the perturbative result has been
  convolved with a Lorentzian of width $\eta$. }
\label{fig:ed4}
\end{center}
\end{figure}

In Figs. \ref{fig:ed1}--\ref{fig:ed4} we compare the dynamical structure
factor calculated for a 16--site chain to the results obtained by our
perturbative approach for several values of momentum.
For $Q=0$, $\pi$ and $T=2J$ (Fig. \ref{fig:ed1}) the agreement of the two
methods is good. The difference at very small freqencies is probably
due to finite--size effects in the exact diagonalization results. The
finite--size effects for the main peak are found to be quite small.

At $Q=\pi/2$ and $T=2J$ (Fig. \ref{fig:ed2}) the agreement of the two
methods is less impressive. The disagreement at high frequencies is
likely due to inaccuracy of the bubble summation in our perturbative
method (we recall that the agreement of our resummation with the exact
result at $T=0$ was worst for $Q=\pi/2$). On the other hand, the exact
diagonalization results are found to still suffer from finite--size
effects at small frequencies. 
\section{Conclusions}
\label{xxzconclusion}
We have calculated the transverse dynamical structure factor of the
XXZ spin chain at finite temperature and applied field. 
The perturbative method we use is accurate at low temperatures and for large
$\Delta \agt 10$. In this case the chain is in the Ising phase and
the excitations are descended from propagating domain walls. The
scattering response is composed of two distinct parts for $\Delta \agt
10$, namely a gapped continuum at $\omega \sim \Delta J$  and the
`Villain mode' at $\omega \sim 0$. Our results pertain to the low
temperature and field dependence of both. Our main observations are
\begin{enumerate}
\item{} The response associated with the gapped two--spinon continuum at $T=0$
narrows in energy as temperature is increased and loses spectral
weight to the emerging low frequency Villain mode.
\item{} The position of the peak (as a function of frequency) of the high frequency (gapped)
  continuum at $T>0$ becomes asymmetric about $Q=\pi/2$ as temperature increases. 
\item{} The thermally activated response at low frequencies $\omega
  \sim 0$ is asymmetric about $Q=\pi/2$. 
\item{} The Villain mode splits into two peaks in a transverse field. 
\end{enumerate}
The main advantage of our method compared to previous theoretical
approaches lies in the fact that it is not restricted to
asymptotically large values of $\Delta$ and treats the Villain mode
and the gapped response in a unified way, which allows us to determine
the ratio of spectral weights between these two features.
Our results are in qualitative agreement with inelastic neutron
scattering experiments.\cite{nagler,nagler2,oosawa,braun} It would be
interesting to perform quantitative comparisons with polarized neutron
scattering data.

\begin{acknowledgments}
We are grateful to Alan Tennant for getting us interested in this
problem and to Isaac P\'{e}rez--Castillo for his collaboration on
the early part of this work and for many helpful discussions
since. In particular we thank him for providing efficient code to
generate the exact solution curves at $T=0$. We thank I. Affleck,
S.E. Nagler and A.M. Tsvelik for useful discussions.
 This work was supported by the EPSRC under grant EP/D050952/1. 
\end{acknowledgments}
\appendix
\section{Direct Jordan--Wigner Transformation of the Hamiltonian}
\label{sec:direct}
We may write the Hamiltonian (\ref{eqnxxzham}) as $H=H_0+H_1$, where
\bea
H_0&=&\frac{J}{4} \sum_n \Delta \sigma_n^z \sigma_{n+1}^z +
\sigma_{n}^y \sigma_{n+1}^y+\frac{h}{2}\sum_n\sigma_n^x\ ,\nn
H_1&=&\frac{J}{4} \sum_n \sigma_{n}^x \sigma_{n+1}^x .
\eea
$H_0$ can then be expressed as a quadratic form in spinless fermions by
means of the Jordan--Wigner transformation
\bea
\sigma^x_n&=&c^\dagger_nc_n-1\ ,\nn
\frac{\sigma_n^z-i\sigma_n^y}{2}&=&-c_n^\dagger e^{-i \pi \sum_{j<n} c^\dagger_j c_j}.
\eea
The full Hamiltonian takes the form
\bea
H&=&\sum_n\frac{J}{4}(\Delta+1)\left[c^\dagger_nc_{n+1}+{\rm
    h.c.}\right]+\frac{h-J}{2}c^\dagger_nc_n\nn
&&+\frac{J}{4}\sum_n(\Delta-1)\left[c^\dagger_nc^\dagger_{n+1}+{\rm
    h.c.}\right]\nn
&&+\frac{J}{4}\sum_nc^\dagger_nc_nc^\dagger_{n+1}c_{n+1}.
\eea
The terms quadratic in fermions can be diagonalized by a Bogoliubov
transformation (\ref{bogtrafo}). The resulting free--fermion dispersion
is different from (\ref{eqnxxzH2}). Expressing the interaction part in
terms of Bogoliubov fermions and normal ordering self--consistently
leads to a theory of the same structure as the one derived in section
\ref{secxxztrans}. In fact we expect it to be identical, but we have
not verified this.
\section{One Vertex Diagrams}
\label{apponevertex}
In this section we list the first order contributions to the transverse response and discuss their divergent behaviour.
\subsection{`Connected Bubble' diagrams}
\label{secrpadiags}
The contributions to $L^k_i\Pi_{ij}^{kq}(\omega,Q)R^q_j$ given by connecting two bubbles are (the box vertices in the diagrams indicate the inclusion of the external factors, $L^k_i$ and $R^q_j$):
\begin{align}
  & \quad
  \parbox{30mm}{\begin{fmfgraph*}(75,40)
      \fmfleft{v1} \fmfright{v3}
      \fmf{fermion,left=0.7}{v2,v1}
      \fmf{fermion,right=0.7}{v2,v1}
      \fmf{fermion,left=0.7}{v3,v2}
      \fmf{fermion,right=0.7}{v3,v2}
      \fmfdot{v2}
      \fmfv{decor.shape=square,decor.filled=empty,decor.size=2thick}{v1,v3}
  \end{fmfgraph*}}  \nonumber \\ \nn
  =& \frac{1}{N^2} \sum_{k,q} \sin(\gamma_k)\sin(\gamma_q) V_2(-k,k+Q,q,-q-Q)\nonumber \\ &\qquad \times \ocp{k}{-}{-} \ocp{q}{-}{-}
\label{bubble1}
\end{align}

\begin{align}
  & \quad
  \parbox{30mm}{\begin{fmfgraph*}(75,40)
      \fmfleft{v1} \fmfright{v3}
      \fmf{fermion,left=0.7}{v1,v2}
      \fmf{fermion,right=0.7}{v1,v2}
      \fmf{fermion,left=0.7}{v2,v3}
      \fmf{fermion,right=0.7}{v2,v3}
      \fmfdot{v2}
      \fmfv{decor.shape=square,decor.filled=empty,decor.size=2thick}{v1,v3}
  \end{fmfgraph*}}  \nonumber \\ \nn
  =& \frac{1}{N^2} \sum_{k,q}\sin(\gamma_k)\sin(\gamma_q) V_2(q,-q-Q,-k,k+Q)  \nonumber \\ &\qquad \times \ocp{k}{+}{+} \ocp{q}{+}{+}
\label{bubble2}
\end{align}

\begin{align}
  & \quad
  \parbox{30mm}{\begin{fmfgraph*}(75,40)
      \fmfleft{v1} \fmfright{v3}
      \fmf{fermion,left=0.7}{v2,v1}
      \fmf{fermion,left=0.7}{v1,v2}
      \fmf{fermion,left=0.7}{v3,v2}
      \fmf{fermion,left=0.7}{v2,v3}
      \fmfdot{v2}
      \fmfv{decor.shape=square,decor.filled=empty,decor.size=2thick}{v1,v3}
  \end{fmfgraph*}}  \nonumber \\ \nn
  =& \frac{4}{N^2} \sum_{k,q} \cos(\gamma_k)\cos(\gamma_q)  V_2(k+Q,q,-k,-q-Q) \nonumber \\
 &\qquad \times \ocm{k}{+}{-} \ocm{q}{+}{-}
\label{bubble3}
\end{align}

\begin{align}
  & \quad
  \parbox{40mm}{\begin{fmfgraph*}(75,40)
      \fmfleft{v1} \fmfright{v3}
      \fmf{fermion,right=0.7}{v1,v2}
      \fmf{fermion,left=0.7}{v1,v2}
      \fmf{fermion,left=0.7}{v3,v2}
      \fmf{fermion,right=0.7}{v3,v2}
      \fmfdot{v2}
      \fmfv{decor.shape=square,decor.filled=empty,decor.size=2thick}{v1,v3}
  \end{fmfgraph*}}  \nonumber \\ \nn
  =& \frac{6}{N^2} \sum_{k,q}\sin(\gamma_k)\sin(\gamma_q) V_0(-k,k+Q,q,-q-Q) \nonumber \\ &\qquad \times \ocp{k}{+}{+} \ocp{q}{-}{-}
\label{bubble4}
\end{align}
\begin{align}
  & \quad
  \parbox{40mm}{\begin{fmfgraph*}(75,40)
      \fmfleft{v1} \fmfright{v3}
      \fmf{fermion,left=0.7}{v2,v1}
      \fmf{fermion,right=0.7}{v2,v1}
      \fmf{fermion,left=0.7}{v2,v3}
      \fmf{fermion,right=0.7}{v2,v3}
      \fmfdot{v2}
      \fmfv{decor.shape=square,decor.filled=empty,decor.size=2thick}{v1,v3}
  \end{fmfgraph*}}  \nonumber \\ \nn
  =& \frac{6}{N^2} \sum_{k,q}\sin(\gamma_k)\sin(\gamma_q) V_0(-k,k+Q,q,-q-Q) \nonumber \\ &\qquad \times \ocp{k}{-}{-} \ocp{q}{+}{+}
\label{bubble5}
 \end{align}
\begin{align}
  & \quad
  \parbox{40mm}{\begin{fmfgraph*}(75,40)
      \fmfleft{v1} \fmfright{v3}
      \fmf{fermion,left=0.7}{v2,v1}
      \fmf{fermion,left=0.7}{v1,v2}
      \fmf{fermion,right=0.7}{v2,v3}
      \fmf{fermion,left=0.7}{v2,v3}
      \fmfdot{v2}
      \fmfv{decor.shape=square,decor.filled=empty,decor.size=2thick}{v1,v3}
  \end{fmfgraph*}}  \nonumber \\ \nn
  =& \frac{3}{N^2} \sum_{k,q}\cos(\gamma_k)\sin(\gamma_q) i V_1(k,q+Q,-q,-k-Q) \nonumber \\ &\qquad \times \ocm{k}{+}{-} \ocp{q}{+}{+}
\label{bubble6}
  \end{align}
\begin{align}
   & \quad
  \parbox{40mm}{\begin{fmfgraph*}(75,40)
      \fmfleft{v1} \fmfright{v3}
      \fmf{fermion,left=0.7}{v2,v1}
      \fmf{fermion,right=0.7}{v2,v1}
      \fmf{fermion,right=0.7}{v3,v2}
      \fmf{fermion,right=0.7}{v2,v3}
      \fmfdot{v2}
      \fmfv{decor.shape=square,decor.filled=empty,decor.size=2thick}{v1,v3}
  \end{fmfgraph*}}  \nonumber \\ \nn
  =& -\frac{3}{N^2} \sum_{k,q} \sin(\gamma_k)\cos(\gamma_q) i  V_1(-q,q+Q,-k-Q,k) \nonumber \\ &\qquad \times \ocp{k}{-}{-} \ocm{q}{-}{+}
\label{bubble7}
\end{align}
\begin{align}
  & \quad
  \parbox{40mm}{\begin{fmfgraph*}(75,40)
      \fmfleft{v1} \fmfright{v3}
      \fmf{fermion,left=0.7}{v2,v1}
      \fmf{fermion,left=0.7}{v1,v2}
      \fmf{fermion,left=0.7,label=$q+Q$}{v3,v2}
      \fmf{fermion,right=0.7,label=$-q$}{v3,v2}
      \fmfdot{v2}
      \fmfv{decor.shape=square,decor.filled=empty,decor.size=2thick}{v1,v3}
  \end{fmfgraph*}}  \nonumber \\ \nn
  =& \frac{3}{N^2} \sum_{k,q}\cos(\gamma_k)\sin(\gamma_q)i V_1^*(k+Q,-k,q,-q-Q) \nonumber \\ &\qquad \times \ocm{k}{+}{-} \ocp{q}{-}{-}
\label{bubble8}
\end{align}
\begin{align}
  & \quad
  \parbox{40mm}{\begin{fmfgraph*}(75,40)
      \fmfleft{v1} \fmfright{v3}
      \fmf{fermion,right=0.7}{v1,v2}
      \fmf{fermion,left=0.7}{v1,v2}
      \fmf{fermion,right=0.7}{v3,v2}
      \fmf{fermion,right=0.7}{v2,v3}
      \fmfdot{v2}
      \fmfv{decor.shape=square,decor.filled=empty,decor.size=2thick}{v1,v3}
  \end{fmfgraph*}}  \nonumber \\ \nn
  =& -\frac{3}{N^2} \sum_{k,q} \sin(\gamma_k)\cos(\gamma_q) i V_1^*(-q-Q,-k,k+Q,q) \nonumber \\ &\qquad \times \ocp{k}{+}{+} \ocm{q}{-}{+}
\label{bubble9}
\end{align}
\subsection{`Tadpole' type diagrams}
These contributions consist of bubbles in which one of the propagators features a tadpole type self energy interaction:
\begin{align}
& \quad
  \left[ \;\;\; \parbox{25mm}{\begin{fmfgraph*}(60,30)
      \fmfleft{v1} \fmfright{v3}
      \fmfforce{(.5w,0.9h)}{v2}
      \fmf{plain,tension=1.4,right=0}{v2,v2}
      \fmf{fermion,left=0.5}{v3,v1}
      \fmf{fermion,right=0.25}{v2,v1}
      \fmf{fermion,right=0.25}{v3,v2}
      \fmfdot{v2}
      \fmfv{decor.shape=square,decor.filled=empty,decor.size=2thick}{v1,v3}
      \end{fmfgraph*}}\!\!\!
      +
      \;\;\;   \parbox{25mm}{\begin{fmfgraph*}(60,30)
      \fmfleft{v1} \fmfright{v3}
      \fmfforce{(.5w,0.1h)}{v2}
      \fmf{plain,tension=1.4,right=0}{v2,v2}
      \fmf{fermion,right=0.5}{v3,v1}
      \fmf{fermion,left=0.25}{v2,v1}
      \fmf{fermion,left=0.25}{v3,v2}
      \fmfdot{v2}
      \fmfv{decor.shape=square,decor.filled=empty,decor.size=2thick}{v1,v3}
  \end{fmfgraph*}}\!\!\!
  \right]  \nonumber \\  \nonumber \\
  =& \frac{8}{N^2} \sum_{k,q} (\cos(2\gamma_k)-1)\frac{n_q V_2(-k,q,-q,k) }{i\omega_n-\epsilon_k-\epsilon_{k+Q}} \nonumber \\
   & \times \left[ \frac{n_k+n_{k+Q}-1}{i\omega_n-\epsilon_k-\epsilon_{k+Q}}- \beta n_k(1-n_k) \right]
\label{tadpole1}
\end{align}
\begin{align}
\nn
& \quad
  \left[ \;\;\; \parbox{25mm}{\begin{fmfgraph*}(60,30)
      \fmfleft{v1} \fmfright{v3}
      \fmfforce{(.5w,0.9h)}{v2}
      \fmf{plain,tension=1.4,right=180}{v2,v2}
      \fmf{fermion,right=0.5}{v1,v3}
      \fmf{fermion,left=0.25}{v1,v2}
      \fmf{fermion,left=0.25}{v2,v3}
      \fmfdot{v2}
      \fmfv{decor.shape=square,decor.filled=empty,decor.size=2thick}{v1,v3}
  \end{fmfgraph*}}\!\!\!
      +
      \;\;\; \parbox{25mm}{\begin{fmfgraph*}(60,30)
      \fmfleft{v1} \fmfright{v3}
      \fmfforce{(.5w,0.1h)}{v2}
      \fmf{plain,tension=1.4,right=180}{v2,v2}
      \fmf{fermion,left=0.5}{v1,v3}
      \fmf{fermion,right=0.25}{v1,v2}
      \fmf{fermion,right=0.25}{v2,v3}
      \fmfdot{v2}
      \fmfv{decor.shape=square,decor.filled=empty,decor.size=2thick}{v1,v3}     
  \end{fmfgraph*}}\!\!\!
  \right] 
  \nonumber \\  \nonumber \\
  =& -\frac{2}{N^2} \sum_{k,q} (\cos(2\gamma_k)-1) \frac{n_q  V_2(k,q,-q,-k)}{i\omega_n+\epsilon_k+\epsilon_{k+Q}}\nonumber \\
   &\times \left[ \frac{n_k+n_{k+Q}-1}{i\omega_n+\epsilon_k+\epsilon_{k+Q}}+ \beta n_k(1-n_k) \right]
\label{tadpole2}
\end{align}
\begin{align}
 & \quad
  \left[ \;\;\;  \parbox{25mm}{\begin{fmfgraph*}(60,30)
      \fmfleft{v1} \fmfright{v3}
      \fmfforce{(.5w,0.9h)}{v2}
      \fmf{fermion,left=0.5}{v3,v1}
      \fmf{fermion,left=0.25}{v1,v2}
      \fmf{fermion,left=0.25}{v2,v3}
      \fmf{plain,tension=1.4,right=180}{v2,v2}
      \fmfdot{v2}
      \fmfv{decor.shape=square,decor.filled=empty,decor.size=2thick}{v1,v3}
  \end{fmfgraph*}}  \!\!\!
      +
      \;\;\;   \parbox{25mm}{\begin{fmfgraph*}(60,30)
      \fmfleft{v1} \fmfright{v3}
      \fmfforce{(.5w,0.1h)}{v2}
      \fmf{fermion,left=0.5}{v1,v3}
      \fmf{fermion,left=0.25}{v2,v1}
      \fmf{fermion,left=0.25}{v3,v2}
      \fmf{plain,tension=1.4,right=0}{v2,v2}
      \fmfdot{v2}
      \fmfv{decor.shape=square,decor.filled=empty,decor.size=2thick}{v1,v3}
  \end{fmfgraph*}}\!\!\!
  \right] \nonumber \\  \nonumber \\
  =& \frac{2}{N^2} \sum_{k,q} (\cos(2\gamma_k)+1) \frac{n_q V_2(k,q,-q,-k)}{i\omega_n+\epsilon_k-\epsilon_{k+Q}} \nonumber \\
  &\times \left[ \frac{n_k-n_{k+Q}}{i\omega_n+\epsilon_k-\epsilon_{k+Q}}+ \beta n_k(1-n_k) \right]\nn
-& \frac{2}{N^2} \sum_{k,q} (\cos(2\gamma_k)+1) \frac{n_q V_2(k+Q,q,-q,-k-Q)}{i\omega_n+\epsilon_k-\epsilon_{k+Q}} \nonumber \\
  & \times\left[ \frac{n_k-n_{k+Q}}{i\omega_n+\epsilon_k-\epsilon_{k+Q}}+ \beta n_{k+Q}(1-n_{k+Q}) \right]
\label{tadpole3}
\end{align}
\begin{align}
 & \quad
  \left[ \;\;\; \parbox{25mm}{\begin{fmfgraph*}(60,30)
      \fmfleft{v1} \fmfright{v3}
      \fmfforce{(.5w,0.9h)}{v2}
      \fmf{plain,tension=2,right=0}{v2,v2}
      \fmf{fermion,left=0.5}{v3,v1}
      \fmf{fermion,left=0.25}{v1,v2}
      \fmf{fermion,right=0.25}{v3,v2}
      \fmfdot{v2}
      \fmfv{decor.shape=square,decor.filled=empty,decor.size=2thick}{v1,v3}
      \end{fmfgraph*}}\!\!\!
  \right]  \nonumber \\
  =& -\frac{3}{N^2} \sum_{k,q} \sin(2\gamma_k)i V_1(q,-q,-k,k) \ocs{k}{+}{-} \nonumber \\ &\quad \times \left[\ocp{k}{-}{-}+\frac{2n(\epsilon_k)-1}{2\epsilon_k}\right]
\label{tadpole4}
\end{align}
\begin{align}
 & \quad
  \left[ \;\;\; \parbox{25mm}{\begin{fmfgraph*}(60,30)
      \fmfleft{v1} \fmfright{v3}
      \fmfforce{(.5w,0.1h)}{v2}
      \fmf{plain,tension=2,right=180}{v2,v2}
      \fmf{fermion,left=0.5}{v1,v3}
      \fmf{fermion,right=0.25}{v1,v2}
      \fmf{fermion,left=0.25}{v3,v2}
      \fmfdot{v2}
      \fmfv{decor.shape=square,decor.filled=empty,decor.size=2thick}{v1,v3}
  \end{fmfgraph*}}\!\!\!
  \right]
  \nonumber \\
  =&- \frac{3}{N^2} \sum_{k,q} \sin(2\gamma_k)i V_1(q,-q,-k,k)  \ocs{k}{+}{+}\nonumber \\ &\quad \times\left[\ocm{k}{-}{+}+\frac{2n(\epsilon_k)-1}{2\epsilon_k}\right]
\label{tadpole5}
\end{align} 
\begin{align}
 & \quad
  \left[ \;\;\;  \parbox{25mm}{\begin{fmfgraph*}(60,30)
      \fmfleft{v1} \fmfright{v3}
      \fmfforce{(.5w,0.9h)}{v2}
      \fmf{fermion,left=0.5}{v3,v1}
      \fmf{fermion,right=0.25}{v2,v1}
      \fmf{fermion,left=0.25}{v2,v3}
      \fmf{plain,tension=2,right=90}{v2,v2}
      \fmfdot{v2}
      \fmfv{decor.shape=square,decor.filled=empty,decor.size=2thick}{v1,v3}
  \end{fmfgraph*}}  \!\!\!
  \right] \nonumber \\
  =&\frac{3}{N^2} \sum_{k,q} \sin(2\gamma_k)i V_1(q,-q,-k,k) \ocs{k}{-}{-}\nonumber \\ &\quad \times \left[\ocm{k}{+}{-}-\frac{2n(\epsilon_k)-1}{2\epsilon_k}\right] \label{tadpole6}\end{align}
 \begin{align}
 & \quad
  \left[ \;\;\;  \parbox{25mm}{\begin{fmfgraph*}(60,30)
      \fmfleft{v1} \fmfright{v3}
      \fmfforce{(.5w,0.1h)}{v2}
      \fmf{fermion,left=0.5}{v1,v3}
      \fmf{fermion,left=0.25}{v2,v1}
      \fmf{fermion,right=0.25}{v2,v3}
      \fmf{plain,tension=2,right=-90}{v2,v2}
      \fmfdot{v2}
      \fmfv{decor.shape=square,decor.filled=empty,decor.size=2thick}{v1,v3}
  \end{fmfgraph*}} \!\!\!
  \right] 
 \nonumber \\
  =& -\frac{3}{N^2} \sum_{k,q} \sin(2\gamma_k)i V_1(q,-q,-k,k)  \ocs{k}{-}{+} \nonumber \\ &\quad \times\left[\ocp{k}{+}{+}-\frac{2n(\epsilon_k)-1}{2\epsilon_k}\right]
\label{tadpole7}\end{align}
One expects that, just as at zeroth order, these diagrams will have
divergences at certain energies. Inspection of the pole structure of the connected bubble diagrams (\ref{bubble1}--\ref{bubble9}) suggests that they will have stronger divergences than the zeroth order because they feature products of poles. Numerical results support this assumption for Eqs. (\ref{bubble1},\ref{bubble2}). On the other hand, it is clear that Eqs. (\ref{bubble4}--\ref{bubble9}) will be very small (for large $\Delta$) because they contain products of terms that, individually, are only significant for different discrete regions in $\omega$.
The behaviour of Eq. (\ref{bubble3}) is subtler. This diagram gives the most significant first order contribution to the response at $\omega \sim 0$. However it does not contain a stronger divergence than its zeroth order equivalent. The reason is as follows: consider the sum
\begin{align}
\sum_k\frac{I(k,Q)}{\omega+\epsilon_{k}-\epsilon_{k+Q}+i\eta}
\end{align}
with $I(k,Q)$ an analytic function of $k,Q$. As shown at zeroth order, because the dispersion $\epsilon_k$ is bounded, the imaginary part of the sum (as $\eta \to 0$) will in turn be bounded. In general there is a divergence at the threshold $\omega=\mathrm{max} (\epsilon_{k}-\epsilon_{k+Q})$ of the corresponding response. Naively one then expects that in the double sum
\begin{align}
\sum_{k,q}\frac{I(k,Q)I(q,Q)I'(k,q,Q)}{(\omega+\epsilon_{k}-\epsilon_{k+Q}+i\eta)((\omega+\epsilon_{k}-\epsilon_{k+Q}+i\eta))}
\end{align}
with $I'(k,q,Q)$ analytic, the maximum contribution will occur for $k=q$ and $\omega=\mathrm{max} (\epsilon_{k}-\epsilon_{k+Q})$.
For Eq. (\ref{bubble3}) the function $V_2(k+Q,k,-k,-k-Q)$ vanishes at the threshold $\omega=\max (\epsilon_{k}-\epsilon_{k+Q})$. This means that the divergence is in fact substantially weaker than at zeroth order. This behaviour does not persist when self--energy corrections to the propagator are included. This is because the self--energy corrections shift the thresholds of the response.
Stronger than leading order divergences are also found in Eqs. (\ref{tadpole1}--\ref{tadpole7}). To take them into account the single particle propagator must be resummed using a Dyson equation, as in Sec. \ref{secxxzsp}.
\section{Matrix Structure of the Bubble Summation }
\label{secrpaproof}
In this appendix we show that summing all bubble diagrams results in
the integral equation (\ref{inteq1}) for the matrix $\Pi^{\rm RPA}$.
The proof follows by induction. 
Our starting point is expressions (\ref{chiPi}) and (\ref{eqnchiU}) for
the dynamical susceptibility. The $n^{\rm th}$ order contribution to
the matrix $\Pi$ is by definition
\be
-\bigl\langle T_\tau
 \boldsymbol{X}(\tau,Q|k)\boldsymbol{X}^\dagger(0,Q|q)U^{(n)}
 \bigr\rangle \ ,
\label{Arpa1}
\ee
where
\be
U^{(n)}= \frac{(-1)^{n}}{n!}\prod_{m=1}^{n} \int_0^\beta  d\tau_m H'_4(\tau_m)\ .
\ee
Out of all possible contractions in (\ref{Arpa1}) we want to select
only those that give rise to bubble diagrams. We denote their
contribution by $\boldsymbol{\Pi}^{\rm RPA}_{(n)}$. We wish to show
that
\be
\boldsymbol{\Pi}^{\rm RPA}_{(n)}(\tau,Q|k,q)=
\Bigl(\boldsymbol{K}\circ\boldsymbol{K}\circ\dots
\circ\boldsymbol{K}\circ
\boldsymbol{\Pi}^0\Bigr)(\tau,Q|k,q),
\label{Arpa2}
\ee
where 
$\boldsymbol{K}$ is defined as
\be
K_{\alpha\beta}(\tau,Q|k,q)=\sum_{k'}\Pi^0_{\alpha\gamma}(\tau,Q|k,k')V_{\gamma\beta}(Q|k',q),
\ee
and $\circ$ denotes a convolution and simultaneous matrix multiplication
\begin{widetext}
\be
\left(\boldsymbol{K}\circ\boldsymbol{K}\right)_{\alpha\beta}(\tau,Q|k,q)=
\frac{1}{N}\sum_{k'}\int d\tau_1
K_{\alpha\gamma}(\tau-\tau_1,Q|k,k')\ K_{\gamma\beta}(\tau_1,Q|k',q).
\ee
\end{widetext}
Fourier transforming (\ref{Arpa2}) then gives
\be
\boldsymbol{\Pi}^{\rm RPA}_{(n)}(i\omega_n,Q|k,q)=
\Bigl(\boldsymbol{K}*\boldsymbol{K}*\dots
*\boldsymbol{K}*\boldsymbol{\Pi}^0\Bigr)(i\omega_n,Q|k,q),
\ee
where the convolution $*$ is defined as
\bea
\left(\boldsymbol{K}*\boldsymbol{K}
\right)_{\alpha\beta}(i\omega_n,Q|k,q)&=&
\frac{1}{N}\sum_{k'}K_{\alpha\gamma}(i\omega_n,Q|k,k')\nn
&&\times\ K_{\gamma\beta}
(i\omega_n,Q|k',q). 
\eea
Summing over $n$ then leads to
\be
\boldsymbol{\Pi}^{\rm RPA}(i\omega_n,Q|k,q)=
\sum_{n=0}^\infty\boldsymbol{\Pi}^{\rm RPA}_{(n)}(i\omega_n,Q|k,q).
\ee
This can be written as
\be
\boldsymbol{\Pi}^{\rm RPA}=\sum_{n=0}^\infty
\boldsymbol{K}^n*\boldsymbol{\Pi}^0
= \left(\boldsymbol{I}-\boldsymbol{K}\right)^{-1}*\boldsymbol{\Pi}^0,
\ee
where
\begin{align}
\boldsymbol{K}^n=\overbrace{\boldsymbol{K}*\boldsymbol{K}\dots*\boldsymbol{K}}^{n}
\end{align}.

The basic identity underlying the inductive proof of (\ref{Arpa2}) is
\begin{align}
\contraction{}{\boldsymbol{X}}{(\tau,Q|k)}{U}
\boldsymbol{X}(\tau,Q|k)U^{(n)}\bigg|_{\rm RPA}
=\left(\boldsymbol{K}\circ\boldsymbol{K}\circ\dots \circ\boldsymbol{K}\circ\boldsymbol{X}\right)(\tau,Q|k),
\label{Arpa3}
\end{align}
where the contraction notation indicates that only contractions compatible with our
RPA-like summation have been carried out.
Eqn (\ref{Arpa2}) is clearly a direct consequence of (\ref{Arpa3}). We
now prove (\ref{Arpa3}) by induction. For $n=1$ we prove by a lengthy
but straightforward calculation that
\begin{align}
\contraction{}{\boldsymbol{X}}{(\tau,Q|k)}{U}
\boldsymbol{X}(\tau,Q|k)U^{(1)}\bigg|_{\rm RPA}
=\left(\boldsymbol{K}\circ\boldsymbol{X}\right)(\tau,Q|k).
\label{Arpa4}
\end{align}
The induction step is then straightforward. We have
\be
\contraction{}{\boldsymbol{X}}{(\tau,Q|k)}{U}
\boldsymbol{X}(\tau,Q|k)U^{(n+1)}\bigg|_{\rm RPA}\!\!\!=\!-\int d\tau_{n+1}
\contraction{}{\boldsymbol{Y}}{}{H'_4}
\boldsymbol{Y}H'_4(\tau_{n+1})\bigg|_{\rm RPA},
\label{Arpa5}
\ee
where 
\be
\boldsymbol{Y}=
\contraction{}{\boldsymbol{X}}{(\tau,Q|k)}{U}
\boldsymbol{X}(\tau,Q|k)U^{(n)}\bigg|_{\rm
  RPA}.
\label{Arpa6}
\ee
The combinatorial factor $n+1$ cancels exactly against the $1/(n+1)$
in the definition of $U^{(n+1)}$.
Now using the induction assumption (\ref{Arpa3}) in (\ref{Arpa6}), the final
contraction in (\ref{Arpa5}) reduces to the induction start
$n=1$, thus establishing the validity of (\ref{Arpa3}) for $n+1$.
\section{Elements of $\boldsymbol{\Pi}_S$}
\label{appelementspis}
Introducing the definitions
\be
B_{--}(k,k+Q) = -\frac{n(E_k)+n(E_{k+Q})-1}{i\omega_n-E_k-E_{k+Q}},
\ee
\be
B_{++}(k,k+Q) = -\frac{1-n(E_k)-n(E_{k+Q})}{i\omega_n+E_k+E_{k+Q}},
\ee
\be
B_{-+}(k,k+Q) = -\frac{n(E_k)-n(E_{k+Q})}{i\omega_n-E_k+E_{k+Q}},
\ee
\be
B_{+-}(k,k+Q) = -\frac{n(E_{k+Q})-n(E_k)}{i\omega_n+E_k-E_{k+Q}},
\ee
the explicit elements of the matrix, Eq. (\ref{PiS}), are
\begin{widetext}
\bea
{\Pi}^S_{11}(i\omega_n,Q|k,k')&=&\sum_{\sigma,\sigma'=\pm}
Z^\sigma_kZ^{\sigma'}_{k+Q}\ B_{\sigma\sigma'}(k,k+Q)
\bigl(\delta_{k,k'}-\delta_{k,-k'-Q}\bigr)\ ,\\
{\Pi}^S_{12}(i\omega_n,Q|k,k')&=&\sum_{\sigma,\sigma'=\pm}
\sigma \lambda_{k'} Z^{\sigma'}_{k'+Q}\ B_{\sigma\sigma'}(k',k'+Q)\bigl(\delta_{k,k'}
-\delta_{k,-k'-Q}\bigr),\\
{\Pi}^S_{13}(i\omega_n,Q|k,k')&=&\sum_{\sigma,\sigma'=\pm}
\sigma\sigma'\lambda_{k+Q}\lambda_{k}\ B_{\sigma\sigma'}(k,k+Q)
\bigl(\delta_{k,k'}-\delta_{k,-k'-Q}\bigr)\ ,
\eea
\bea
{\Pi}^S_{21}(i\omega_n,Q|k,k')&=&-\sum_{\sigma,\sigma'=\pm}
\sigma \lambda_{k} Z^{\sigma'}_{k+Q}\ B_{\sigma\sigma'}(k,k+Q)
\bigl(\delta_{k,k'}-\delta_{k,-k'-Q}\bigr)\ ,\\
{\Pi}^S_{22}(i\omega_n,Q|k,k')&=&\sum_{\sigma,\sigma'=\pm}\bigl(
Z^{\sigma'}_{k+Q}Z^{-\sigma}_{k}
\delta_{k,k'}
+\sigma\sigma'\lambda_{k}\lambda_{k+Q}
\delta_{k,-k'-Q} \bigr)\ B_{\sigma\sigma'}(k,k+Q),\\
{\Pi}^S_{23}(i\omega_n,Q|k,k')&=&\sum_{\sigma,\sigma'=\pm}\sigma'
\lambda_{k+Q} Z^{-\sigma}_{k}\ B_{\sigma\sigma'}(k,k+Q)
\bigl(\delta_{k,k'}-\delta_{k,-k'-Q}\bigr)\ ,
\eea
\bea
{\Pi}^S_{31}(i\omega_n,Q|k,k')&=&\sum_{\sigma,\sigma'=\pm}
\sigma\sigma'\lambda_{k+Q}\lambda_{k}\ B_{\sigma\sigma'}(k,k+Q)
\bigl(\delta_{k,k'}-\delta_{k,-k'-Q}\bigr)\ ,\\
{\Pi}^S_{32}(i\omega_n,Q|k,k')&=&-\sum_{\sigma,\sigma'=\pm}
\sigma'\lambda_{k'+Q}Z^{-\sigma}_{k'}\ B_{\sigma\sigma'}(k',k'+Q)\bigl(\delta_{k,k'}
-\delta_{k,-k'-Q}\bigr)\ ,\\
{\Pi}^S_{33}(i\omega_n,Q|k,k')&=&\sum_{\sigma,\sigma'=\pm}
Z^{-\sigma}_{k}Z^{-\sigma'}_{k+Q}\ B_{\sigma\sigma'}(k,k+Q)
\bigl(\delta_{k,k'}-\delta_{k,-k'-Q}\bigr)\ .
\eea
\end{widetext}

\section{Further Contributions to the XXZ Spin Chain Response}
\label{app2loopsp}
The calculation described in the main text leads to a response that features sharp thresholds even at finite temperature. Physically one expects these thresholds to be absent at $T\ne 0$. The diagrams we consider are incapable of capturing this effect because they include poles that only depend on two single particle energies. This limits the response at positive frequencies to the regions ${\rm min}(E_k+E_{q}) \le \omega \le {\rm max}(E_k+E_{q})$ and $0 \le \omega \le {\rm max}(E_k-E_{q})$, for general $k,q$. Coupling to decay channels involving more than two particles should alleviate this problem.

One can evaluate the two--loop self energy correction to the propagator. This includes diagrams of the form
\begin{align}
  \parbox{20mm}{\begin{fmfgraph*}(50,30)
      \fmfleft{v1} \fmfright{v4}
      \fmf{fermion,tension=4}{v2,v1}
      \fmf{fermion,tension=4}{v4,v3}
      \fmf{fermion,left=0.9}{v3,v2}
      \fmf{fermion,right=0.9}{v3,v2}
      \fmf{fermion}{v3,v2}
      \fmfdot{v2}
      \fmfdot{v3}
  \end{fmfgraph*}},\;
  \parbox{20mm}{\begin{fmfgraph*}(50,30)
      \fmfleft{v1} \fmfright{v4}
      \fmf{fermion,tension=4}{v2,v1}
      \fmf{fermion,tension=4}{v4,v3}
      \fmf{fermion,left=0.9}{v3,v2}
      \fmf{fermion,left=0.9}{v2,v3}
      \fmf{fermion}{v3,v2}
      \fmfdot{v2}
      \fmfdot{v3}
  \end{fmfgraph*}},\;
  \parbox{20mm}{\begin{fmfgraph*}(50,30)
      \fmfleft{v1} \fmfright{v4}
      \fmf{fermion,tension=4}{v2,v1}
      \fmf{fermion,tension=4}{v3,v4}
      \fmf{fermion,left=0.9}{v3,v2}
      \fmf{fermion,left=0.9}{v2,v3}
      \fmf{fermion}{v2,v3}
      \fmfdot{v2}
      \fmfdot{v3}
  \end{fmfgraph*}},\; \hdots
\label{eqn2loopsp}
\end{align}
In analogy with the one--loop calculation, we define a self energy matrix $\boldsymbol{\Sigma}_{2}$. As
before the transferred frequency and momentum are labelled by $ik_n$
and $k$ respectively. 
Defining $r=k+p+q$, the elements of the matrix $\Sigma_{2}^{ij}$ are
\begin{widetext}
\bea
\Sigma_{2}^{12} &=&
-\frac{24}{N^2}\sum_{p,q}\tilde{V}_1(k,-r,p,q)\tilde{V}_0(k,-r,p,q)
\frac{n(\epsilon_q)+n(\epsilon_r)-1}{ik_n+\epsilon_p+\epsilon_q+\epsilon_r}
(n(\epsilon_p)-n_B(\epsilon_q+\epsilon_r)-1)\nn
&&-\frac{24}{N^2}\sum_{p,q}\tilde{V}_1(k,-r,p,q)\tilde{V}_0(-q,-p,r,-k)
\frac{n(\epsilon_q)+n(\epsilon_r)-1}{ik_n-\epsilon_p-\epsilon_q-\epsilon_r}
(n(\epsilon_p)-n_B(\epsilon_q+\epsilon_r)-1)\nn
&& +\frac{12}{N^2}\sum_{p,q}\tilde{V}^\star_1(r,-p,-q,-k)\tilde{V}_2(r,-k,-p,-q)
\frac{n(\epsilon_q)-n(\epsilon_r)}{ik_n+\epsilon_p+\epsilon_q-\epsilon_r}
(n(\epsilon_p)+n_B(\epsilon_r-\epsilon_q)) \nn
&& +\frac{12}{N^2}\sum_{p,q}\tilde{V}^\star_1(r,-p,-q,-k)\tilde{V}_2(k,-r,p,q)
\frac{n(\epsilon_q)-n(\epsilon_r)}{ik_n-\epsilon_p-\epsilon_q+\epsilon_r}
(n(\epsilon_p)+n_B(\epsilon_r-\epsilon_q)).
\eea
\bea
\Sigma_{2}^{11}&=&\frac{96}{N^2}\sum_{p,q}\bigl(\tilde{V}_0(k,-r,p,q)\bigr)^2
\frac{n(\epsilon_q)+n(\epsilon_r)-1}{ik_n+\epsilon_p+\epsilon_q+\epsilon_r}
(n(\epsilon_p)-n_B(\epsilon_q+\epsilon_r)-1) \nn
&&+\frac{6}{N^2}\sum_{p,q}\lvert\tilde{V}_1(k,-r,p,q)\rvert^2
\frac{n(\epsilon_q)+n(\epsilon_r)-1}{ik_n-\epsilon_p-\epsilon_q-\epsilon_r}
(n(\epsilon_p)-n_B(\epsilon_q+\epsilon_r)-1)\nn 
&&+\frac{18}{N^2}\sum_{p,q}\lvert\tilde{V}_1(r,-q,-p,-k)\rvert^2
\frac{n(\epsilon_q)-n(\epsilon_r)}{ik_n+\epsilon_p+\epsilon_q-\epsilon_r}
(n(\epsilon_p)+n_B(\epsilon_r-\epsilon_q)) \nn
&&+\frac{8}{N^2}\sum_{p,q}\bigl(\tilde{V}_2(k,-r,p,q)\bigr)^2
\frac{n(\epsilon_q)-n(\epsilon_r)}{ik_n-\epsilon_p-\epsilon_q+\epsilon_r}
(n(\epsilon_p)+n_B(\epsilon_r-\epsilon_q)),
\eea
\end{widetext}
Here we have used the boson occupation factor, $n_B(\epsilon)=1/(\exp(\beta \epsilon)-1)$.
The remaining elements of the two loop self energy are obtained via
\begin{align}
\Sigma_{2}^{22}&=-(\Sigma_{2}^{11})^\star, \\
\Sigma_{2}^{21}&=(\Sigma_{2}^{12})^\star.
\end{align}
These contributions have poles that feature three single particle energies. Some of these contributions will lead to a temperature dependent broadening of the response in the vicinities of $\omega \sim \Delta J$ and $\omega \sim 0$. Including the two--loop self--energy terms should further increase the quality of our approximation. However the extra sum over the loop momentum means that such a computation would be a factor of $N \sim 400$ slower.
\end{fmffile}


\begin{thebibliography}{99}

\bibitem{orbach}
R.~Orbach, {\em Phys. Rev.} {\bf 112}, 309 (1958).

\bibitem{cloizeaux}
J.~des Cloizeaux and M.~Gaudin, {\em J. Math. Phys.} {\bf 7}, 1384 (1966).

\bibitem{yang1}
C.~N. Yang and C.~P. Yang, {\em Phys. Rev.} {\bf 150}, 321 (1966).

\bibitem{korepin}
V. E. Korepin, N. M. Bogoliubov and A. G. Izergin, {\em Quantum Inverse Scattering Method and Correlation Functions}, Cambridge University Press, Cambridge, England (1993).

\bibitem{algra}
H.~A. Algra, L.~J. de~Jongh, H.~W.~J. Bl{\"o}te, W.~J. Huiskamp, and R.~L.
  Carlin, {\em Physica (Utrecht)} {\bf 82B}, 239 (1976).

\bibitem{duxbury}
P.~M. Duxbury, J.~Oitmaa, M.~N. Barber, A.~van~der Bilt, K.~O. Joung, and R.~J.
  Carlin, {\em Phys. Rev. B} {\bf 24}, 5149 (1981).

\bibitem{yoshizawa}
H.~Yoshizawa, G.~Shirane, H.~Shiba, and K.~Hirakawa, {\em Phys. Rev. B} {\bf 28}, 3904 (1983).

\bibitem{nagler}
S.~E. Nagler, W.~J.~L. Buyers, R.~L. Armstrong, and B.~Briat, {\em Phys.\ Rev. Lett.} {\bf 49}, 590 (1982).

\bibitem{nagler2}
S.~E. Nagler, W.~J.~L. Buyers, R.~L. Armstrong, and B.~Briat, {\em Phys.\ Rev. B} {\bf 28}, 3873 (1983).

\bibitem{yoshizawa2}
H.~Yoshizawa, K.~Hirakawa, S.~K. Satija, and G.~Shirane, {\em Phys. Rev. B} {\bf 23}, 2298 (1981).

\bibitem{oosawa}
A.~Oosawa, K.~Kakurai, Y.~Nishiwaki, and T.~Kato, {\em J. Phys. Soc. Jpn.} {\bf 75}, 074719 (2006).

\bibitem{faddeev}
L. D. Faddeev and L. A. Takhtajan, {\em Phys. Lett.} {\bf 85A}, 375 (1981).

\bibitem{braun}
H.-B. Braun, J. Kulda, B. Roessli, D. Visser, K. W. Kr{\"{a}}mer, H.-U. G{\"{u}}del, and P. B{\"{o}}ni, {\em Nat. Phys.} {\bf 1}, 159 (2005).

\bibitem{igor} 
I. Zaliznyak and S. Lee, {\it Magnetic Neutron Scattering} in Modern
Techniques for Characterizing Magnetic Materials, ed. Y. Zhu,
Springer, Heidelberg (2005). 

\bibitem{villain}
J.~Villain, {\em Physica} {\bf 79B}, 1 (1975).

\bibitem{essrmk} 
F. H. L. Essler and R. M. Konik, {\em Phys. Rev. B} {\bf 78}, 100403(R) (2008).

\bibitem{dimerchain} 
A. J. A. James, F. H. L. Essler and R. M. Konik, {\em Phys. Rev. B} {\bf 78}, 094411 (2008).

\bibitem{mikeska} 
H. J. Mikeska and C. Luckmann, Phys. Rev. B {\bf 73}, 184426 (2006).

\bibitem{tennant1}
D. A. Tennant et al., unpublished.

\bibitem{is}
N.~Ishimura and H.~Shiba, {\em Prog. Theor. Phys.} {\bf 63}, 743 (1980).

\bibitem{isaac}
J.-S. Caux, J.~Mossel, and I.~P\'{e}rez-Castillo, {\em J. Stat. Mech.}, P08006 (2008).

\bibitem{jimbobook}
M.~Jimbo and T.~Miwa, {\em Algebraic Analysis of Solvable Lattice Models}, American Mathematical Society, Providence, Rhode Island (1995).

\bibitem{bg}
A.~H. Bougourzi, M.~Karbach, and G.~M{\"{u}}ller, {\em Phys.\ Rev. B} {\bf 57}, 11429 (1988).

\bibitem{thermo}
M. Takahashi, Prog. Theor. Phys. {\bf 46}, 401 (1971);
M. Takahashi and M. Suzuki, Prog. Theor. Phys. {\bf 48}, 2187 (1972);
C. Destri and H.J. de Vega, Phys. Rev. Lett. {\bf 69}, {2313} (1992);
A. Kl\"umper, Z. Phys. {\bf 91}, 507 (1993);
C. Destri and H.J. de Vega, Nucl. Phys. B {\bf 438}, {413} (1995);
K. Fabricius, A. Kl\"umper and B.M. McCoy, Phys. Rev. Lett. {\bf 82},
5365 (1999).

\bibitem{fabsbook}
F.~H.~L.~Essler, H.~Frahm, F.~G{\"o}hmann, A.~Kl{\"u}mper, and V.~E.~Korepin, {\it The One--Dimensional Hubbard Model}, Cambridge
University Press, Cambridge (2005).

\bibitem{takahashibook}
M.~Takahashi, {\em Thermodynamics of One--Dimensional Solvable Models}, Cambridge University Press, Cambridge, England (2005).

\bibitem{johnson1}
J. D. Johnson, {\em J. Appl. Phys.} {\bf 52}, 1991 (1981).

\bibitem{mikeska2} 
H. J. Mikeska, Phys. Rev. B {\bf 12}, 2794 (1975).

\bibitem{kitanine}
N. Kitanine, J. M. Maillet, N. A. Slavnov and V. Terras, {\em Nucl. Phys. B} {\bf 729}, 558  (2005)

\bibitem{gohmann1}
F. G{\"{o}}hmann, A. Kl\"{u}mper and A. Seel, {\em J. Phys. A: Math. Gen.} {\bf 37}, 7625  (2004).

\bibitem{gohmann2}
F. G\"{o}hmann, N. P. Hasenclever and A. Seel, {\em J. Stat. Mech.} P10015 (2005).

\bibitem{sakai}
K. Sakai, {\em J. Phys. A: Math. Theor.} {\bf 40}, 7523 (2007). 

\bibitem{finiteTXXZ}
K. Fabricius, U. L\"ow and J. Stolze, Phys. Rev. B{\bf 55}, 5833
(1997);
S. Grossjohann and W. Brenig, arXiv:0811.1956;
T. Barthel, U. Schollw\"ock and S. White, arXiv:0901.2342.

\bibitem{jw}
P.~Jordan and E.~Wigner, {\em Z. Phys.} {\bf 47}, 631 (1928).

\bibitem{kw}
H. A. Kramers and G. H. Wannier, {\em Physical Review} {\bf 60}, 252 (1941).

\bibitem{gs}
G.~G\'{o}mez-Santos, {\em Phys.\ Rev. B} {\bf 41}, 6788 (1990).

\bibitem{cs2cucl4}
D.V. Dmitriev, V.Y. Krivnov and A.A. Ovchinnikov, Phys. Rev. B
{\bf 65}, 172409 (2002);
D.V. Dmitriev, V.Ya. Krivnov, A.A. Ovchinnikov and A. Langari,
JETP {\bf 95}, 538 (2002);
J.-S. Caux, F.H.L. Essler and U. L\"ow, Phys. Rev. B{\bf 68},
134431 (2003);
F. Capraro and C. Gros, Eur. Phys. J. {\bf B29}, 35 (2002).
R. Hagemans, J.S. Caux and U. L\"ow, Phys. Rev. B{\bf 71}, 014437
(2005). 

\bibitem{kogut}
J.~B. Kogut, {\em Rev. Mod. Phys.} {\bf 51}, 659 (1979).

\end{thebibliography}
\end{document}